\documentclass[a4paper,10pt]{elsarticle}
\usepackage{lineno,hyperref}
\usepackage{amsmath}
\usepackage{amssymb}
\usepackage{booktabs}
\usepackage{graphicx}
\usepackage{xcolor}

\begin{document}

\begin{frontmatter}

\title{Charged Analogues of Isotropic Compact Stars Model with Buchdahl Metric in General Relativity}
\author[mymainaddress]{Amit Kumar Prasad}

\author[mymainaddress]{Jitendra Kumar\corref{mycorrespondingauthor}}
\cortext[mycorrespondingauthor]{Corresponding author}
\ead{jitendark@gmail.com}

\address[mymainaddress]{Department of Mathematics,Central University of Jharkhand,Ranchi-835205 India.}

\begin{abstract}
In this work, we examine a spherically symmetric compact body with isotropic pressure profile, in this context we obtain a new class of exact solutions of Einstein's-Maxwell field equation for compact stars with uniform charged distributions on the basis of Pseudo-spheroidal space time with a particular form of electric field intensity and the metric potential $g_{rr}$. Indicating these two parameters takes into account further examination to be done in deciding unknown constants and depicts the compact strange star candidates likes PSR J1614-2230, 4U 1608-52, SAX J1808.4-3658, 4U 1538-52, SMC X-1, Her X-1 and Cen X-3.By the isotropic Tolman-Oppenhimer-Volkoff(TOV)equation, We explore the equilibrium among hydrostatic, gravitational and electric forces. Then, we analyze the stability of model through adiabatic index($\gamma$) and velocity of sound ($0<\dfrac{dp}{c^2d\rho}<1$). we additionally talk about other physical features of this model like, for example, the pressure, redshift, density,energy conditions and mass-radius of the stars in detail and demonstrate that our results are satisfying all the basic prerequisites of a physically legitimate stellar model. We have seen the measurement of basic physical parameter such as pressure,density,energy and redshift are satisfied the reality condition. All the physical quantities such as density,pressure pressure-density ratio and speed of sound is monotonically decreasing. The outcomes acquired are valuable in exploring the strength of other  compact objects like white dwarfs, gravastars and neutron stars. Finally,we have show that the obtain solutions are compatible with observational data for compact objects.
\end{abstract}
\end{frontmatter}
\section{Introduction}
The first exact solution of the Einstein field equations obtained by Karl Schwarzschild\cite{k}.In a theoretical sense, stars are confined in gas and dust clouds with  non-uniform matter circulation and scattered all through generally cosmic systems.In astronomy, compact objects are typically alluded on the whole to white dwarfs or neutron stars. From the time of Sir Isaac Newton, our comprehension of the idea of gravity has progressed however mysteries in physics still remain.  Einstein’s theory of General Relativity (GR) is one of the best essential speculations of gravity in physics. In spite of the fact that its success, numerous expansions of the first Einstein conditions has been researched to satisfactory present observational information on both cosmological and astrophysical scales. Observational information has been educated us that the universe is experiencing a period of quickened extension.The microscopic structure and properties of a dense matter on phenomenal conditions are necessary to examine for compact object. Also, the high exactness information from Type Ia supernovae \cite{sin}, the cosmic microwave background anisotropies \cite{white}, baryonic acoustic motions \cite{alam} and from gravitational lensing are giving help to stellar. The information appears to show that the universe is directly overwhelmed by two obscure components one pressure less dark matter (DM) and second dark energy (DE).
This is in light of the fact that at such uncommon densities nuclear matter may include nucleons and leptons just as a couple of fascinating segments in their different structures also, stages, for example, mesons, hyperons, baryon resonances similarly as strange quark matter (SQM). Be that as it may, it is as yet impractical to find a far-reaching portrayal of the very dense matter in a firmly cooperating system. So it is valuable to examine a definite composition and the idea of molecule connections in the inside of such object.In context of general theory of relativity, a broadly the pursued course is to indicate an equation of state and afterward solve the Einstein field equations for the study of the composition of a compact star.This is also useful for Tolman-Oppenheimer-Volkoff (TOV) equation (see \cite{35,36}) or the condition of hydro dynamical equilibrium. It is conceivable that anisotropic matter is a significant fixing in numerous astrophysical objects for example, stars, gravastars, and so on. Generally, extensive exertion has been committed to picking up a far reaching comprehension of properties of the anisotropic matte with the expectation of delivering physically suitable models of compact stars. Specifically, compact stars may before long give data about the gravitational connection in an extraordinary gravitational condition. \par
Most of the compact stars divided between a strange star and a normal neutron star. Many authors\cite{VV,VV1,RP} studied that the strange stars possess ultra-strong electric fields on their surfaces. The impact of energy densities on ultra-high electric fields  of compact stars was investigated  in\cite{sray,malheiro,fweber,fweber1,fweber2,rp1}. It additionally has been demonstrated that the electric fields increment the stellar mass by up to $ 30\% $ relying upon the quality of it. As opposed to the strange star the surface electric field on account of neutron star is absent\cite{JV}.Based on these properties, one to observationally recognize quark stars from neutron stars.\par
The important characteristic of many astrophysical objects, like compact stars, gravastars etc.,is isotropic matter.In an extreme gravitational environments, compact stars provide information about the gravitational interaction. The extreme internal density and strong gravity of compact star indicate that the pressure within such objects have two different types of pressures, namely the radial and tangential pressure, and these are to be equal. These information of isotropic matter may producing physically valid models of compact stars. Bowers
and Liang \cite{bowers} have noticed about the structure and evolution of relativistic compact objects in general relativity.They investigated the changes in the gravitational mass and surface redshift by generalisation of the equation of hydrostatic equilibrium and obtained a static spherically symmetric configuration. Ruderman \cite{ruderman} analyzed that at high densities of order  $10^{15} g/cm^3$ nuclear matter transformed in anisotropic in nature. Also, they point out that the radial pressure may not be
equal to the tangential one in massive stellar objects. Based on above physical condition, many contentions have been presented for the presence of anisotropy in star models for example, by the presence of type 3A superfluid \cite{kippenhahn}, various types of phase transitions \cite{AI},mixture of two fluids,the presence of solid core or by other different physical marvels.Also exact solution of Einstein's field is important for study of astrophysical object, because many exact solutions of Einstein's field equations have been found but some of them satisfied all the physical  plausibility conditions. This shows the complexity in getting exact solutions of Einstein's field equations describing physically realizable astrophysical objects. Several workers have  charged stars on spheroidal space-time have been studied by Patel and Kopper\cite{10}, Sharma et al.\cite{11}, Gupta and Kumar\cite{12}, Komatiraj and Maharaj\cite{13}. Many projects are suggested by Ivanov\cite{Ivanov} for constructing charged fluid spheres.Recently Naveen and Bijalwan et al.\cite{Bijalwan}\cite{2011 b} for all $K$  except for $0<K<1$ and J.Kumar et al. \cite{Pratibha,2013,2014,amit,amit1} for $0<K<1$ have been obtained perfect fluid charged analogues models with a specific electric intensity. These information  explicitly demonstrates that the models really compares to strange stars in their mass and and radius. Many authors estimated the masses for the stars. The SAX J1808.4−3658 has a mass of $0.9\pm 0.3M_{\odot}$(Elebert et al.\cite{E}).Abubekerov et al.\cite{V} reported the mass of Her X-1 to be  $0.85\pm 0.15M_{\odot}$.Rawls et al.\cite{R} reported the mass of 4U 1538-52 to be $0.87\pm 0.07M_{\odot}$,Cen X-3 to be $1.49\pm 0.08M_{\odot}$ and SMC X-1 to be $1.29\pm 0.05M_{\odot}$.Demorest et al.\cite{demo} reported the mass of PSR J1614-2230 to be $1.97\pm 0.04M_{\odot}$.Guver et al.\cite{guver} reported the mass of 4U 1608-52 to be $1.74\pm 0.14M_{\odot}$.
   \par
\begin{figure}
 \begin{center}
 \includegraphics[width=6cm]{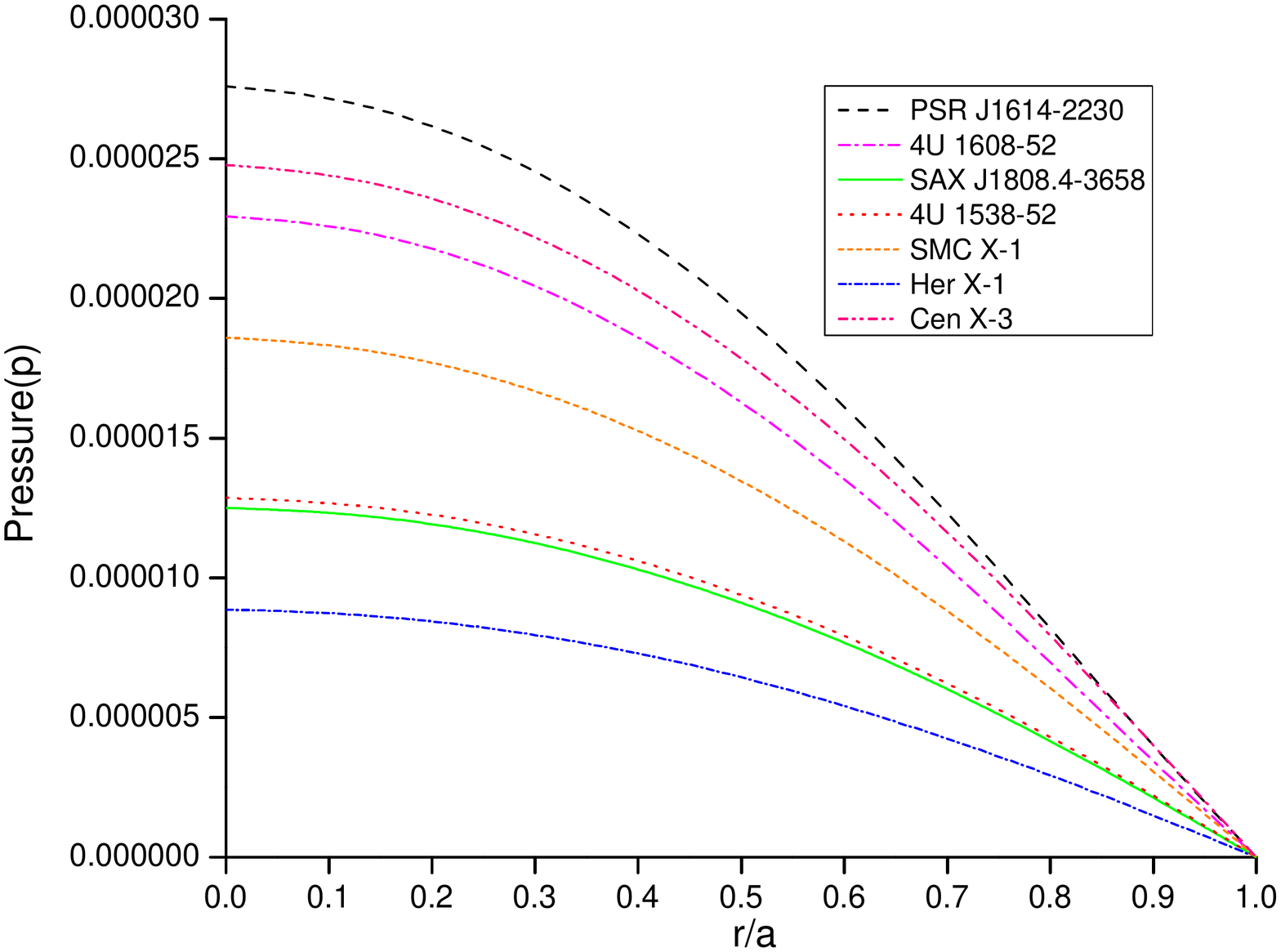} \includegraphics[width=6cm]{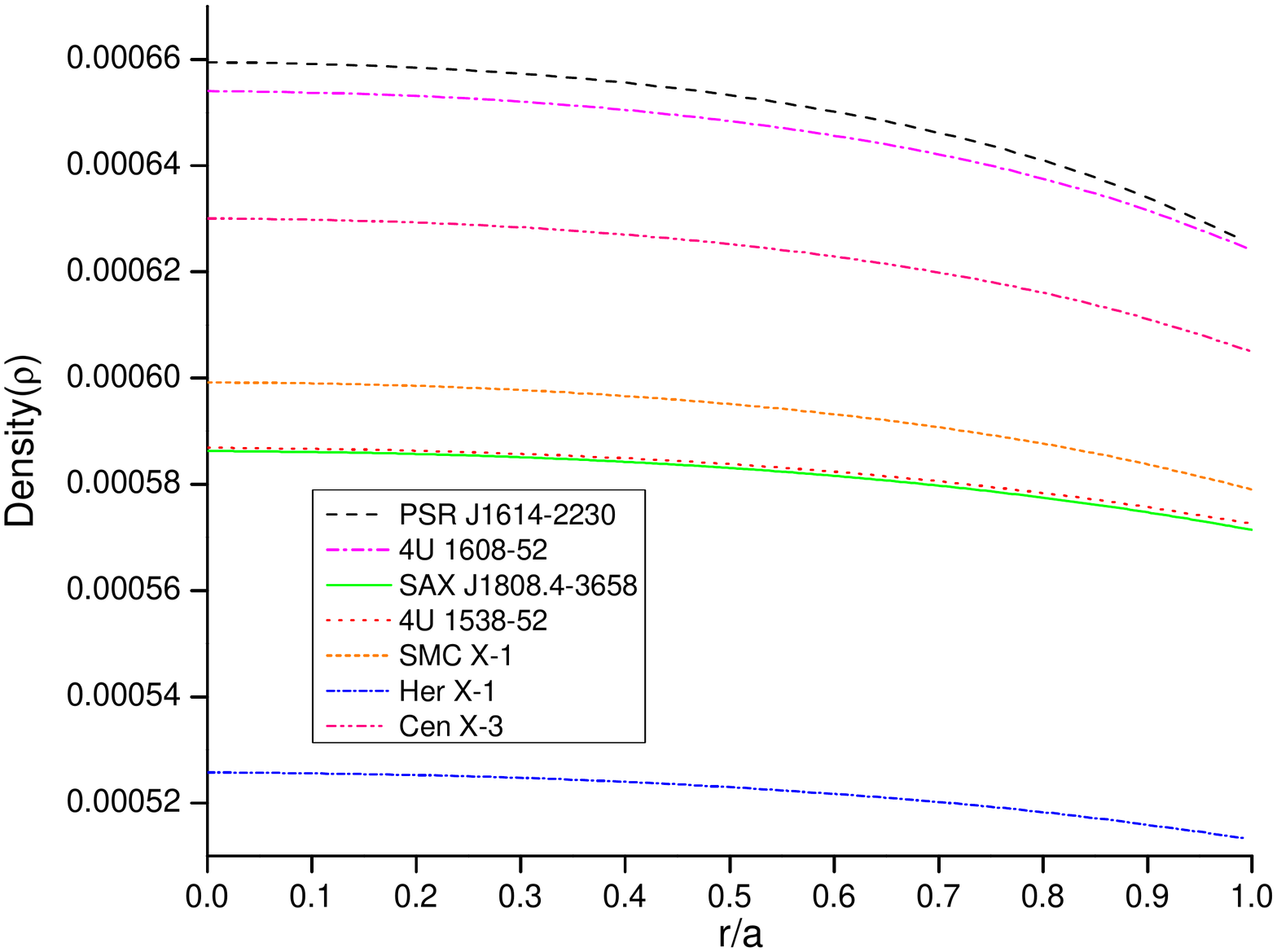}
 \caption{Behavior of pressure(p in $ km^{-2} $ ) and Density ($ \rho $ in $ km^{-2} $ ) vs. fractional radius
 r/a for the compact objects PSR J1614-2230,4U 1608-52, SAX J1808.4-3658, 4U 1538-52,SMC X-1,Her X-1 and Cen X-3. For this figure, we have used the numerical values of physical parameters and constants are as follows: (i) K=0.0000135,b = 0.09,$C=-7.455\times10^{-8}km^{-2}$, $\eta^{2}$=6, M=$01.97M_{\odot}$ and $ a $ = 9.69 km for PSR J1614-2230,(ii)K = 0.0000209, b = 0.0903, $C=-1.1446\times10^{-7}km^{-2}$,$ \eta^{2} $=4.9, M =$01.74M_{\odot}$ , and $ a $ =9.3 km for 4U 1608-52,(iii)K =0.0000319,b=0.09,$C=-1.1564\times10^{-7}km^{-2}$,$ \eta^{2} $=3.9, M =$0.9M_{\odot}$ , and $ a $ =7.951 km for SAX J1808.4-3658,(iv)K = 0.0000296, b = 0.0903, $C=-1.4545\times10^{-7}km^{-2}$,$ \eta^{2} $=4, M =$0.87M_{\odot}$ , and $ a $ =7.866 km for 4U 1538-52,(v) K = 0.000023, b = 0.1,$C=-1.154\times10^{-7}km^{-2}$,$ \eta^{2} $=5, M =$1.29M_{\odot}$ , and $ a $ =8.831 km for SMC X-1,(vi) K = 0.0000277, b = 0.06, $C=-1.2193\times10^{-7}km^{-2}$,$\eta^{2}$=2.9, M = $0.85M_{\odot}$ and $ a $ = 8.1 km for Her X-1,(vii) K = 0.000018, b = 0.09,$C=-9.4971\times10^{-8}km^{-2}$,$ \eta^{2} $=5, M =$1.49M_{\odot}$ , and $ a $ =9.178 km for Cen X-3}\label{f1}
 \end{center}
 \end{figure}

 In the present problem we have constructed a charged
fluid sphere starting with a specific metric potential $g_{44}$ and generalized charge intensity. Delgaty-Lake\cite{8} and Pant et al.\cite{9} have proposed that the physically valid solution in curvature coordinates, the following conditions should be satisfied
\begin{enumerate}
\item At the boundary $ r=a $, pressure $p$ should be zero. 
\item $ c^{2}\rho$ should always be grater than $ p $ within the range $ 0\leq r\leq a .$
\item  The pressure gradient $ dp/dr $ should be negative for $ 0<r \leq a $, i.e,$ (dp/dr)_{r=0}=0 $ and $ (d^{2}p/dr^{2})_{r=0}<0.$ 
\item The density gradient $ d\rho/dr $ should also be negative for $ 0<r \leq a $, i.e, $ (d\rho/dr)_{r=0}=0 $ and $ (d^{2}\rho/dr^{2})_{r=0}<0$ .\\
These two conditions state that the pressure and density should be decreasing towards the surface see Fig.\ref{f1}.
\item The velocity of sound should not exceed the speed of light i.e, $(dp/c^{2}d \rho)^{1/2}<1 $.
\item The adiabatic constant $ \gamma=\left(\left(\dfrac{c^{2}\rho+p}{p}\right)\left(\dfrac{dp}{c^{2}d\rho}\right)\right)>4/3, $ is condition for stability of a fluid sphere.
\item The surface redshift $ Za $ should be positive and finite.
\end{enumerate}
These features, positive density and positive pressure are the most important features characterizing a star. The task is now to check the well-behaved geometry and capability of describing realistic stars, we plot Figs. \ref{f1}-\ref{f6}. Our stellar model is depending on the different values of $K,\eta, b,\textrm{and}\, Ca^2 $. Such analytical representations have been performed by using recent measurements of mass and radius of neutron stars,PSR J1614-2230, 4U 1608-52, SAX J1808.4-3658, 4U 1538-52, SMC X-1, Her X-1 and Cen X-3.
\section{Einstein field equations}
Let us consider the static spherically symmetric metric in curvature coordinates \begin{eqnarray}
 ds^{2}=-e^{\lambda(r)}dr^{2} -r^{2}(d\theta^{2}+\sin^{2}\theta d\phi^{2})+e^{\nu(r)}dt^{2}\label{1}
\end{eqnarray} 
where $\lambda(r)$ and $\nu(r)$ are satisfy the Einstein-Maxwell equation for charged fluid distribution  \begin{eqnarray}R^{i}_{j}-\dfrac{1}{2}R\delta^{i}_{j}=-\kappa \big[(c^{2}\rho+p)\nu^{i}\nu_{j}-p\delta^{i}_{j}+\dfrac{1}{4\pi}(-F^{im}F_{jm}+\dfrac{1}{4}\delta^{i}_{j}F_{mn}F^{mn})\big]\label{2}
\end{eqnarray}
with $ \kappa=\dfrac{8\pi G}{c^{4}}$ while $\rho,p,\nu^{i}$ denote matter density,fluid pressure and the unit time-like flow vector respectively and $F_{ij}$ denote the skew symmetric electromagnetic field tensor. \par
In view of (\ref{1}) the equation (\ref{2}) reduce to (Landau and Lifshitz\cite{Landau}) \begin{eqnarray}
\dfrac{\nu'}{r}e^{-\lambda}-\dfrac{(1-e^{-\lambda})}{r^{2}}=\kappa p-\dfrac{q^{2}}{r^{4}}\label{3} \\
\bigg(\dfrac{\nu''}{2}-\dfrac{\lambda'\nu'}{4}+\dfrac{\nu^{'2}}{4}+\dfrac{\nu'-\lambda'}{2r}\bigg)e^{-\lambda}=\kappa p+\dfrac{q^{2}}{r^{4}}\label{4}\\
\dfrac{\lambda'}{r}e^{-\lambda}+\dfrac{(1-e^{-\lambda})}{r^{2}}=\kappa c^{2}\rho+\dfrac{q^{2}}{r^{4}}\label{5}
\end{eqnarray}
where (') prime denotes the differentiation with respect to $r$ and
  \begin{eqnarray} 
 q(r)=4\pi\int_{0}^{r}\sigma r^{2}e^{\lambda/2}dr=r^{2}\sqrt{-F_{14}F^{14}}=r^{2}F^{41}e^{(\lambda+\nu)/2}\label{6} \end{eqnarray}
 $ q(r) $ represents the total charge contained within the sphere of radius $ r $ in view of (\ref{1}). Additionally,the component $ F_{14}\neq 0 $. On the further side of the pressure free interface  $ \textquoteleft r=a \textquoteright $ the charged fluid sphere is expected to join with the Reissner-Nordstrom metric:  \begin{eqnarray}
 ds^{2}=-\bigg(1-\dfrac{2M}{r}+\dfrac{e^{2}}{r^{2}}\bigg)^{-1}dr^{2}-r^{2}(d\theta^{2}+\sin^{2}\theta d\phi^{2})+\bigg( 1-\dfrac{2M}{r}+\dfrac{e^{2}}{r^{2}}\bigg)dt^{2}\label{7} 
\end{eqnarray}
 where $ M $ is the gravitational mass of the fluid sphere such that \begin{eqnarray} M=\zeta(a)+\xi(a)\label{8}\end{eqnarray} where \begin{eqnarray}
 \zeta(a)=\dfrac{\kappa}{2}\int_{0}^{a}\rho r^{2}dr,\ \ \ \xi(a)=\int_{0}^{a}r\sigma qe^{\lambda/2}dr,\ \ \ e=q(a)\label{9}
 \end{eqnarray}
 where $ \xi(a) $ is the mass equivalence to electromagnetic energy of distribution, $ \zeta(a) $ is the mass and $\textquoteleft e \textquoteright $ is the total charge interior of the sphere (Florides\cite{Florides}). \par In this model we propose a charged fluid distributions by considering the generalized electric field intensity against\cite{amit}  \begin{eqnarray} \dfrac{q^{2}}{r^{4}}=\dfrac{C^2r^2}{2(1+Cr^2)^2}(f_{1}+f_{2})\label{eq11} \end{eqnarray}
 where $ f_1=\dfrac{Cr^2(4K-1)-(K+2)}{4K(1+Cr^2)}, ~~~~~~
 f_2=-\dfrac{2\eta^2(1-K)}{4K(Y^2\eta^2+Yb)(1+Cr^2)} $ \\
 and the metric potential\\
  \begin{eqnarray} e^{\lambda}=\dfrac{K(1+Cr^2)}{K+Cr^2},~~0<K<1\label{10} \end{eqnarray} 
  where $C, K, \eta $ being constants.\\
  The consistency of the field Eqs. (\ref{3})-(\ref{5}) using Eqs. (\ref{eq11}) and (\ref{10}) yield the
  equation  \\ 
  \begin{eqnarray}
  (1+Y^2)\dfrac{d^2Z}{dY^2}-Y\dfrac{dZ}{dY}-[1-K+K(f_1+f_2)]=0\label{eq12}
  \end{eqnarray}
  where  $ Y=\sqrt{\dfrac{K+x}{1-K}},~~~~~ Cr^{2}=x  $ and $ e^{\nu}=Z^{2} $.\\
  The expression for energy density and pressure can be had from (\ref{3}), (\ref{5}), (\ref{eq11})
  and (\ref{10}) as\\
 \begin{eqnarray}
\dfrac{(K+x)}{\sqrt{xC}K(1+x)}\dfrac{2Z'}{Z}+\dfrac{(1-K)}{K(1+x)}+\dfrac{x}{2(1+x)^2}(f_1+f_2)= \dfrac{\kappa p}{C}\label{12}\\
\dfrac{(K-1)(3+x)}{K(1+x)^{2}}-\dfrac{x}{2(1+x)^2}(f_1+f_2)=\dfrac{\kappa c^{2}\rho}{C}\label{13}
  \end{eqnarray}
Let  \begin{eqnarray}
 Z=(1+Y^{2})^{1/4}\Phi(Y)\label{14}
 \end{eqnarray}
 Put the values of $ Z $ from equation (\ref{14}) into equation (\ref{eq12}) we get \begin{eqnarray}
\dfrac{d^{2}\Phi}{dY^{2}}+\tau\Phi=0\label{15}
 \end{eqnarray}
 where
$\, \tau=-\dfrac{2\eta^{2}}{(Y^{2}\eta^{2}+Yb)}\label{17}$\\
Hence the solution of the  differential equation (\ref{15}) is \\ 
 $\Phi(Y)=(Y^{2}\eta^{2}+Yb)\dfrac{A\eta^2}{ b^{3}}\bigg[\sin^2\bigg(\arctan\bigg(\sqrt{\dfrac{Y\eta^2}{b}}\bigg)\bigg)-\csc^2\bigg(\arctan\bigg(\sqrt{\dfrac{Y\eta^2}{b}}\bigg)\bigg)-\\~~~~~~~~~~~~~~~~~~~~~~~~~ \log\sin^4\bigg(\arctan\bigg(\sqrt{\dfrac{Y\eta^2}{b}}\bigg)\bigg)\bigg]+B(Y^{2}\eta^{2}+Yb) $\\ \\
 In the equation (\ref{14}) $ Z $ becomes 
 \begin{eqnarray}
 Z=A(1+Y^{2})^{1/4}\Bigg[(Y^{2}\eta^{2}+Yb)\dfrac{\eta^2}{ b^{3}}G(Y)+B(Y^{2}\eta^{2}+Yb)\Bigg]\label{21}
 \end{eqnarray}
 where\\ $ G(Y)=\bigg[\sin^2\bigg(\arctan\bigg(\sqrt{\dfrac{Y\eta^2}{b}}\bigg)\bigg)-\csc^2\bigg(\arctan\bigg(\sqrt{\dfrac{Y\eta^2}{b}}\bigg)\bigg)-\log\sin^4\bigg(\arctan\bigg(\sqrt{\dfrac{Y\eta^2}{b}}\bigg)\bigg)\bigg] $\\ \\
   Now put the equation (\ref{21}) into equations (\ref{12})-(\ref{13}),we get the expressions of density and pressure
\begin{eqnarray}
\dfrac{\kappa c^{2}\rho}{C}=\dfrac{(K-1)(3+x)}{K(1+x)^{2}}-\dfrac{x}{2K(1+x)^{2}}\bigg[K-1+\dfrac{2-3Y^{2}}{4(1+Y^{2})^{2}}+\dfrac{2\eta^{2}(1+Y^{2})}{(Y^{2}\eta^{2}+Yb)}\bigg]\label{22}\\ \nonumber\\\dfrac{\kappa p}{C}=\bigg(\dfrac{2(K+x)}{K(1+x)\sqrt{(1-K)(K+x)}}\bigg)\bigg[\dfrac{E1\times E2+E3\times E4}{E5\times E2}\bigg]-\dfrac{(K-1)}{K(1+x)}+E6\label{23}
\end{eqnarray}   
where,$E1=\left[\dfrac{\left(\dfrac{K+x}{1-K}\right)^{3/2}\eta^{2}+\left(\dfrac{K+x}{1-K}\right)b}{2\left(\dfrac{1+x}{1-K}\right)^{3/4}}+\left(\dfrac{1+x}{1-K}\right)^{1/4}\left(2\sqrt{\dfrac{K+x}{1-K}}\,\eta^{2}+b\right)\right]\\ \\
E2=\dfrac{\eta^2}{b^3}\bigg[\sin^2\bigg(\arctan\bigg(\sqrt{\dfrac{Y\eta^2}{b}}\bigg)\bigg)-\csc^2\bigg(\arctan\bigg(\sqrt{\dfrac{Y\eta^2}{b}}\bigg)\bigg)-\\ \log\sin^4\bigg(\arctan\bigg(\sqrt{\dfrac{Y\eta^2}{b}}\bigg)\bigg)\bigg]+\dfrac{B}{A},~~~~~~~~~~~~~~~~~~~~~~~~~~~~~~~
 E3=\dfrac{\eta^3}{2}\bigg(\dfrac{Y}{b^5}\bigg)^{1/2} \bigg(1+Y^2\bigg)^{1/4}\\ \\
 E4=\sin2\bigg(\arctan\bigg(\sqrt{\dfrac{Y\eta^2}{b}}\bigg)\bigg)+2\csc^2\bigg(\arctan\bigg(\sqrt{\dfrac{Y\eta^2}{b}}\bigg)\bigg)\times \cot\bigg(\arctan\bigg(\sqrt{\dfrac{Y\eta^2}{b}}\bigg)\bigg)-4\cot\bigg(\arctan\bigg(\sqrt{\dfrac{Y\eta^2}{b}}\bigg)\bigg),\\ \\
 E5=\bigg(1+Y^2\bigg)^{1/4}\bigg(Y^2\eta^2+Yb\bigg),~~~~~~~~~~~
 E6=\dfrac{x}{2K(1+x)^{2}}\bigg[K-1+\dfrac{2-3Y^{2}}{4(1+Y^{2})^{2}}+\dfrac{2\eta^{2}(1+Y^{2})}{(Y^{2}\eta^{2}+Yb)}\bigg]$\\
 
The expression of velocity of sound is as follows, 
\begin{eqnarray}
\dfrac{dp}{c^{2}d\rho}=\dfrac{\dfrac{L}{(E5\times E2)^{2}}-\dfrac{2\sqrt{Cx}}{K(1+Y^{2})^{2}}+E7\times E8+E9\times E10} {\bigg(E11-E7\times E8-E9\times E10\bigg)}\label{24}
\end{eqnarray}
where, $ 
L=(E5\times E2)\times M1\times(E1\times E2+E3\times E4)+M7\bigg[(E5\times E2) \bigg((B2+B3)\times E2+E1\times B4+E3\times B6+E4\times B5\bigg)-(E1\times E2+E3\times E4)(B7\times E2+B4\times E5)\bigg]\\ \\
M1=\dfrac{2\sqrt{Cx}(1-2K-x)}{K(1+x)^2\sqrt{(1-K)(K+x)}},~~~~~~~M7=\dfrac{2(K+x)}{K(1+x)\sqrt{(1-K)(K+x)}},\\ 
E8=\bigg[K-1+\dfrac{2-3Y^{2}}{4(1+Y^{2})^{2}}+\dfrac{2\eta^{2}(1+Y^{2})}{(Y^{2}\eta^{2}+Yb)}\bigg],~~~~~~~E10=\dfrac{x}{2K(1+x)^{2}},\\ \\ \\ 
E9=\dfrac{5(K-1)\sqrt{Cx}}{2K(1+x)^{2}}-\dfrac{-2\eta^{2}}{(Y^{2}\eta^{2}+b)}\dfrac{2\sqrt{Cx}}{K(1-K)}-\dfrac{(1+Y^{2})}{K}\dfrac{2\eta^2\sqrt{Cx}(2\eta^{2}+b/Y)}{(1-K)(Y^{2}\eta^{2}+b)^{2}},\\ \\ 
E7=\dfrac{\sqrt{Cx}(1-x)}{(1+x)^3},~~~~~~~~
B2=\dfrac{\sqrt{Cx}}{(1-K)}\left[\dfrac{(1+Y^2)\Bigg(\dfrac{3}{2}Y\eta^2+b\Bigg)-\dfrac{3}{4}Y^2\Bigg(Y\eta^2+b\Bigg)}{(1+Y^2)^{7/4}}\right],\\ \\ B3=\dfrac{2\sqrt{Cx}}{(1-K)(1+Y^2)^{3/4}}\left[\Bigg(\dfrac{1}{2}Y\eta^2+b\Bigg)+(1+Y^2)\dfrac{\eta^2}{Y}\right],~~~~~~E11=\dfrac{2\sqrt{Cx}(1-K)(5-x)}{K(1+x)^{3}}\\ \\
B4=\dfrac{\eta^3\sqrt{Cx}}{2b^{5/2}(1-K)\Bigg(b+Y\eta^2\Bigg)}\times E4,~~~~~~~~~~~
B5=\dfrac{\eta^3\sqrt{Cx}}{4b^{5/2}(1-K)}\left[(1+Y^2)^{-3/4}Y^{1/2}+(1+Y^2)^{1/4}Y^{3/2}\right]\\
B6=\dfrac{\eta\sqrt{Cxb}}{Y^{3/2}(1-K)(b+Y\eta^2)}\Bigg[\cos2\bigg(\arctan\bigg(\sqrt{\dfrac{Y\eta^2}{b}}\bigg)\bigg)-2\csc^2\bigg(\arctan\bigg(\sqrt{\dfrac{Y\eta^2}{b}}\bigg)\bigg)\times \cot^2\bigg(\arctan\bigg(\sqrt{\dfrac{Y\eta^2}{b}}\bigg)\bigg)-2\csc^4\bigg(\arctan\bigg(\sqrt{\dfrac{Y\eta^2}{b}}\bigg)\bigg)+\csc^2\bigg(\arctan\bigg(\sqrt{\dfrac{Y\eta^2}{b}}\bigg)\bigg)\Bigg]\\ \\ 
B7=\dfrac{\sqrt{Cx}}{2(1-K)(1+Y^2)^{3/4}}(Y^2\eta^2+bY)+\dfrac{2(1+Y^2)^{1/4}\sqrt{Cx}}{(1-K)}\Bigg(\eta^2+\dfrac{b}{Y}\Bigg) $\\
Observing the Fig.\ref{f5}(top right),it shows that velocity of sound lies within the proposed range for different compact stars as labeled in figure.
\section{Boundary conditions}
 The charged fluid sphere is expected to join smoothly with the Reissner-Nordstrom metric (\ref{7}).The continuity of $ e^{\lambda},e^{\nu} $ and $ q $ across the boundary  $ r=a $ gives following equations\\ \\ \begin{eqnarray} e^{\lambda}=1-\dfrac{2m(a)}{a}+\dfrac{e^{2}}{a^{2}},\label{25}\\ y^{2}=1-\dfrac{2m(a)}{a}+\dfrac{e^{2}}{a^{2}},\label{26}\\q(a)=e,\label{27}\\p(a)=0.\label{28}
\end{eqnarray}
The conditions (\ref{26}) and (\ref{28}) can be used to compute the values of arbitrary constants $ A/B. $
\section{Tolman-Oppenheimer-Volkoff (TOV) equations}
In the presence of charge, the Tolman-Oppenheimer-Volkoff(TOV) equation \cite{35,36}  is given by
\begin{figure}[]
\begin{center}
\includegraphics[width=5cm]{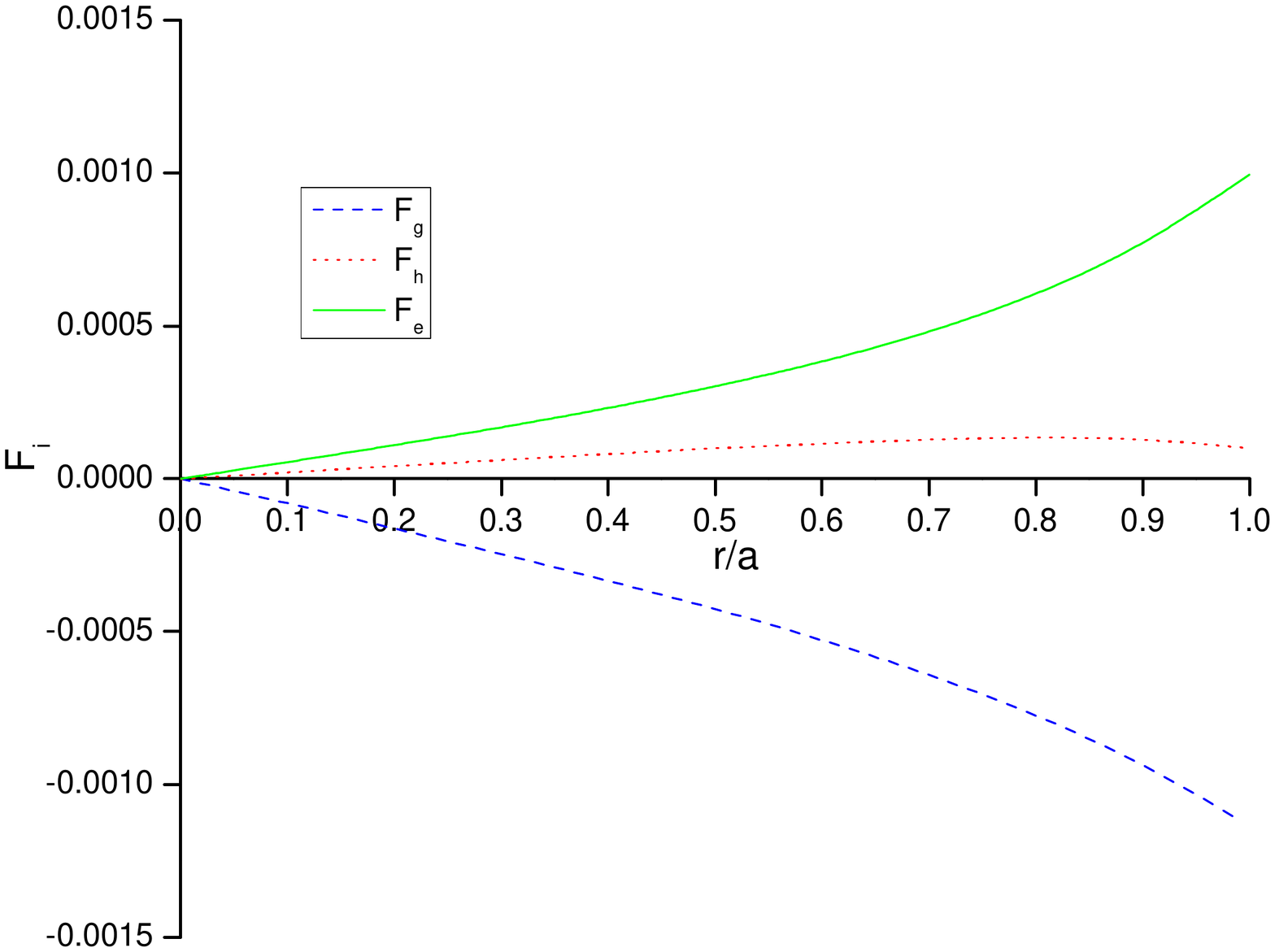}\includegraphics[width=5cm]{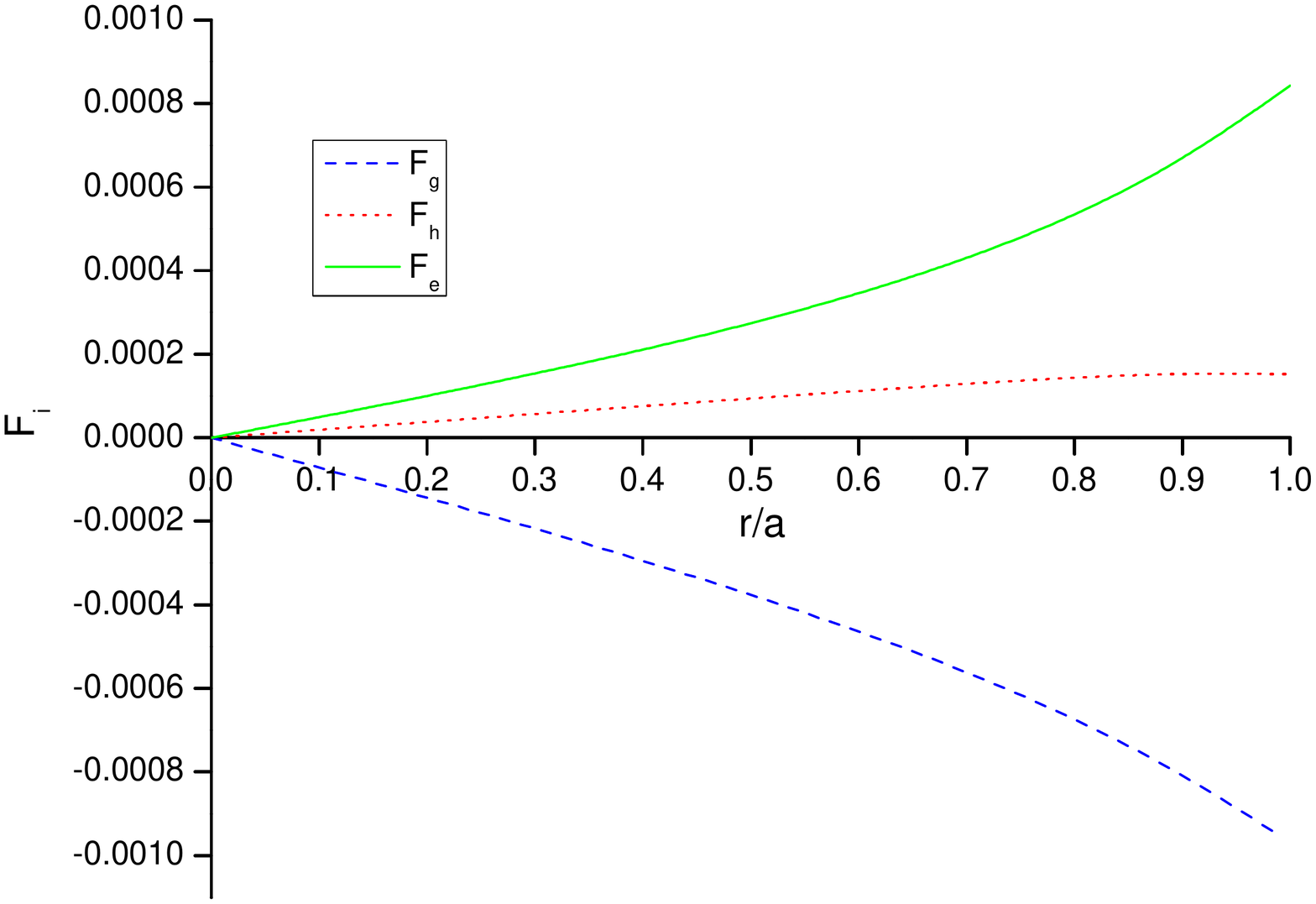}\includegraphics[width=5cm]{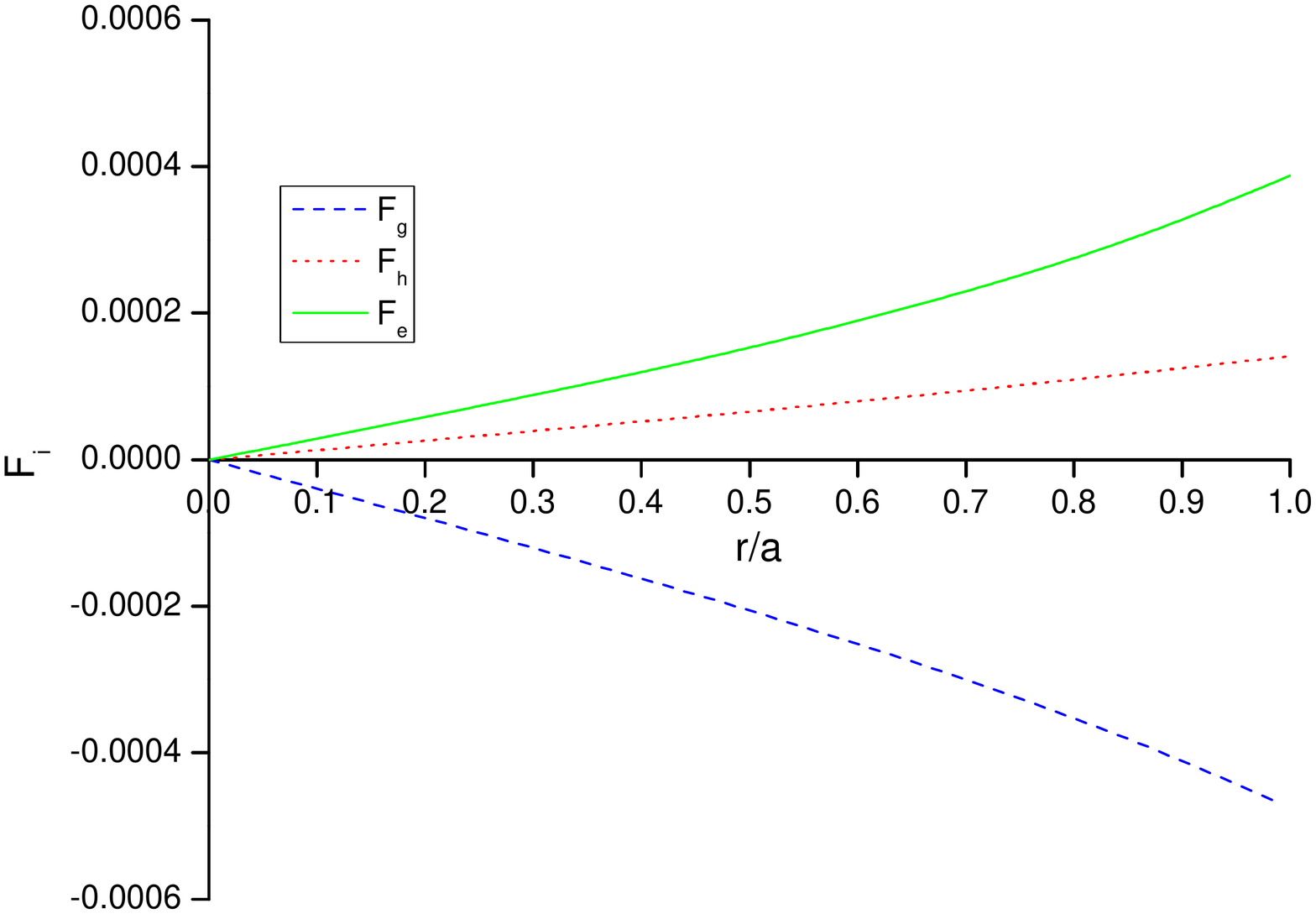}\\\includegraphics[width=5cm]{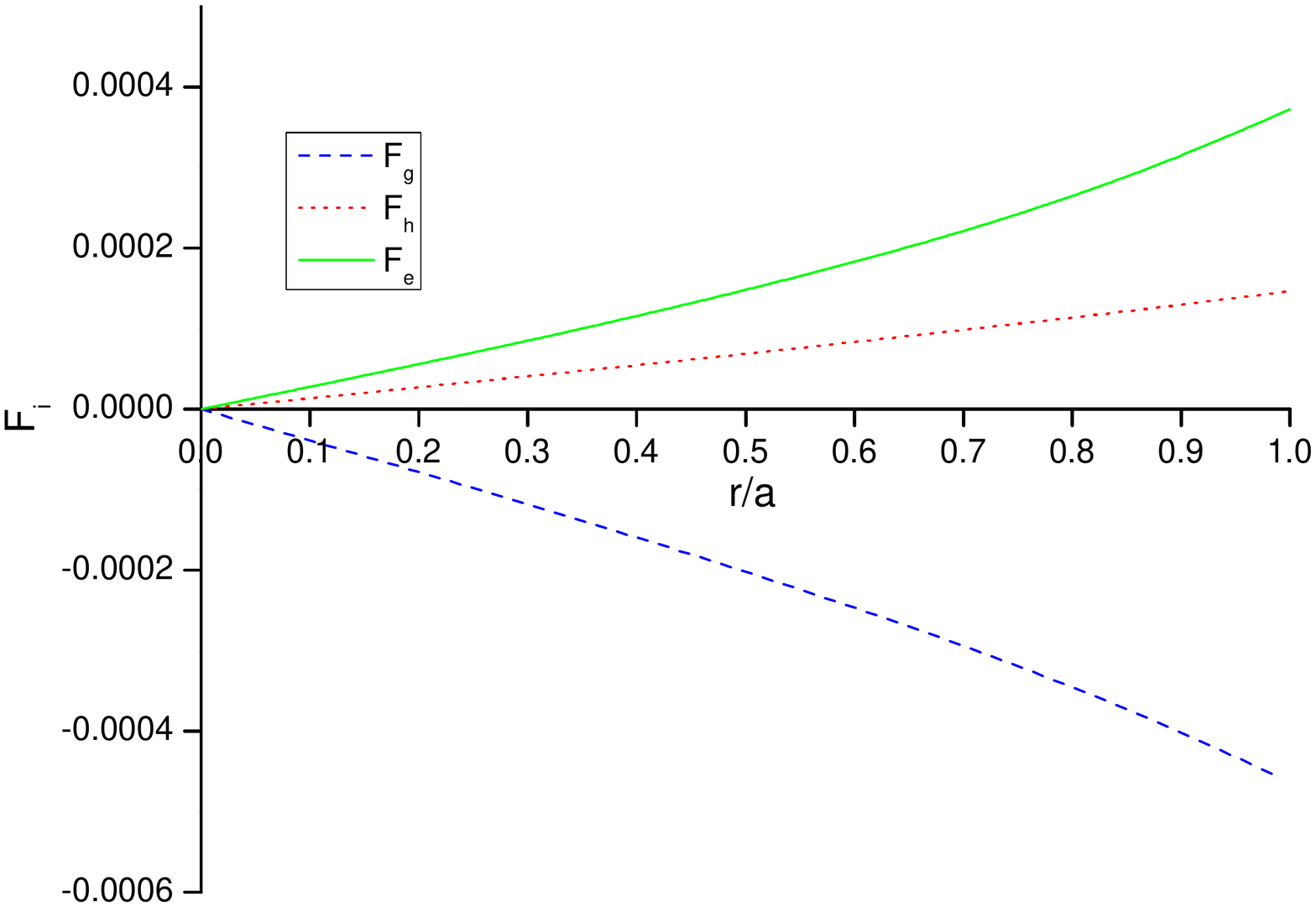}\includegraphics[width=5cm]{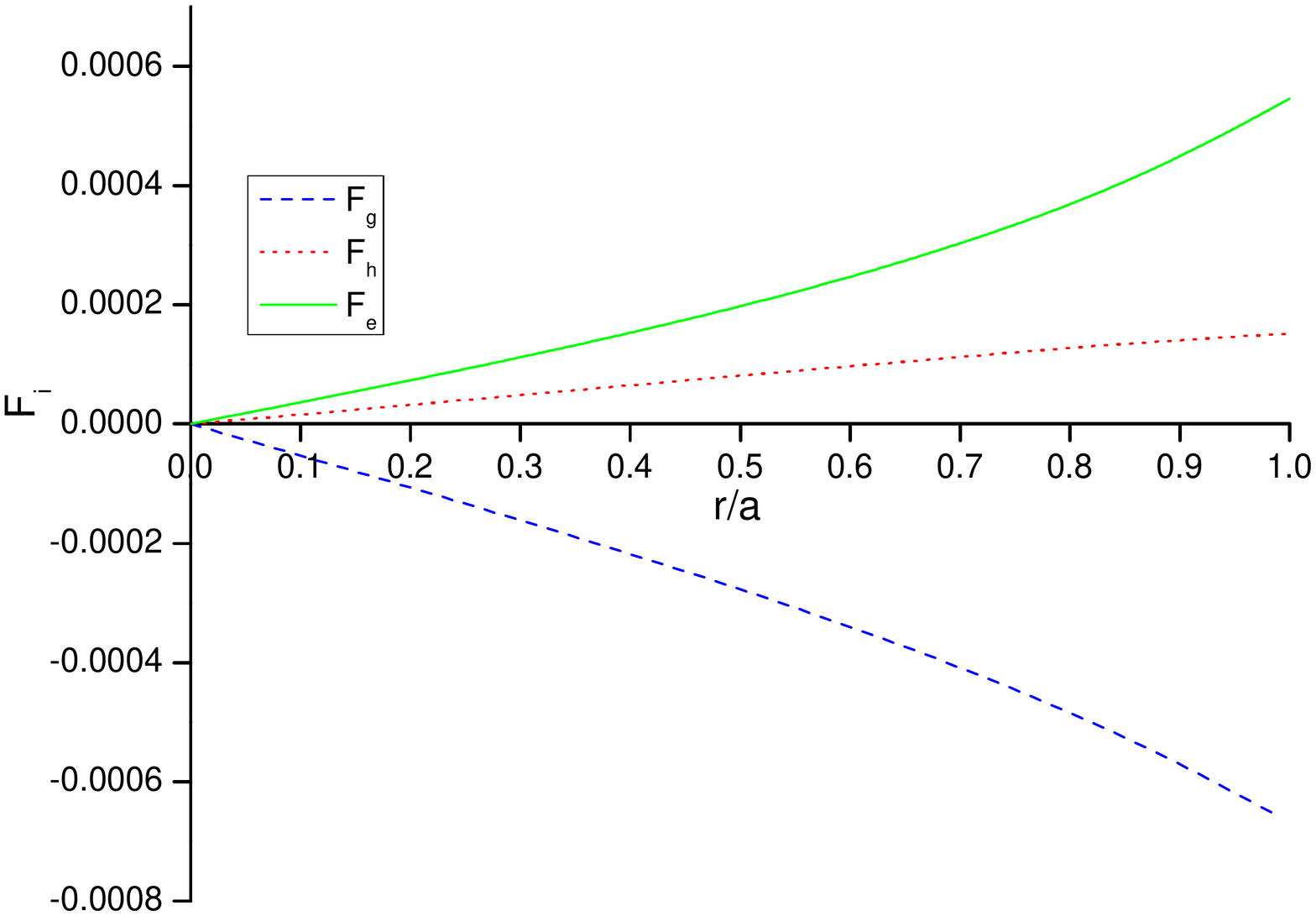}\includegraphics[width=5cm]{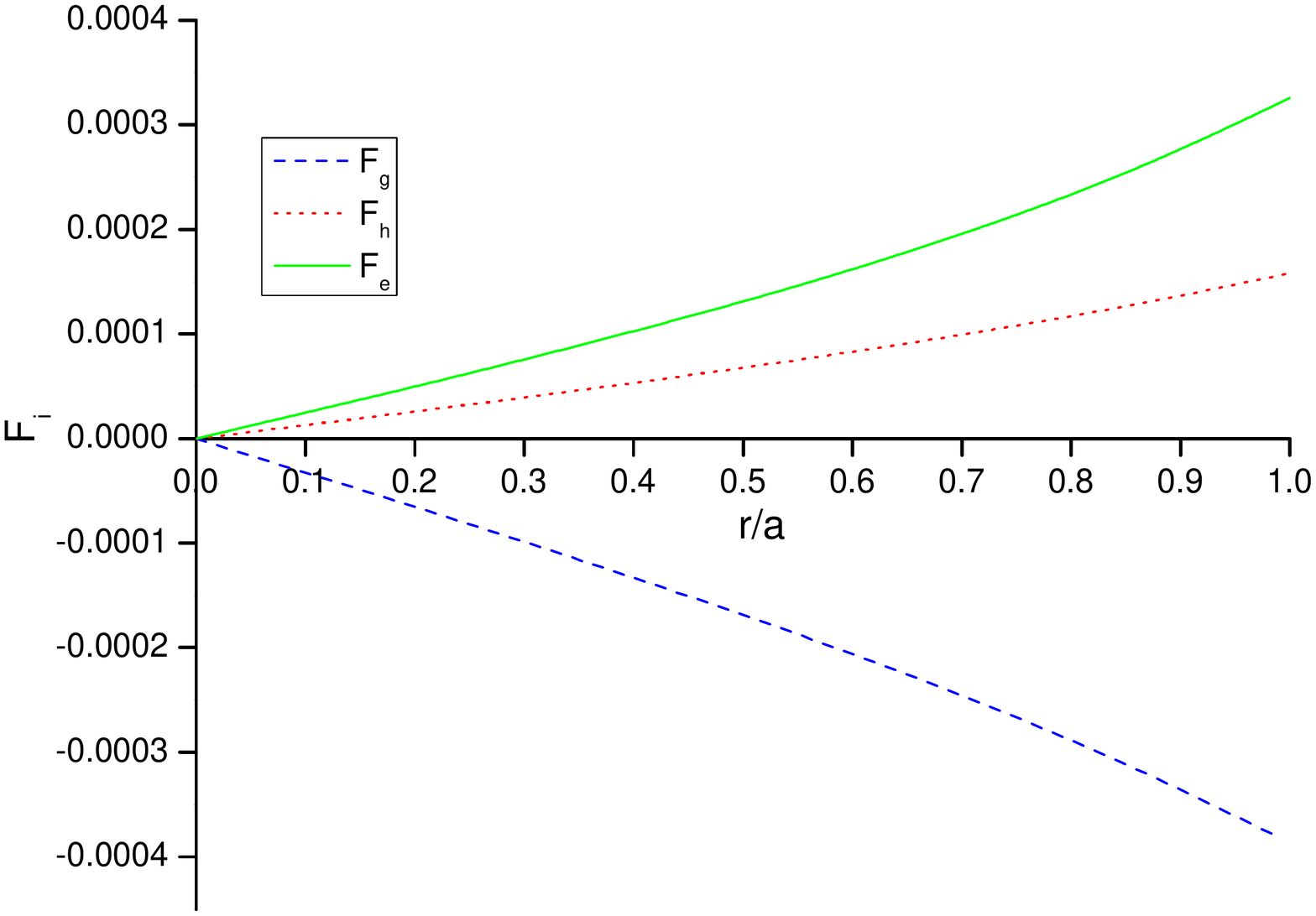}\\\includegraphics[width=5cm]{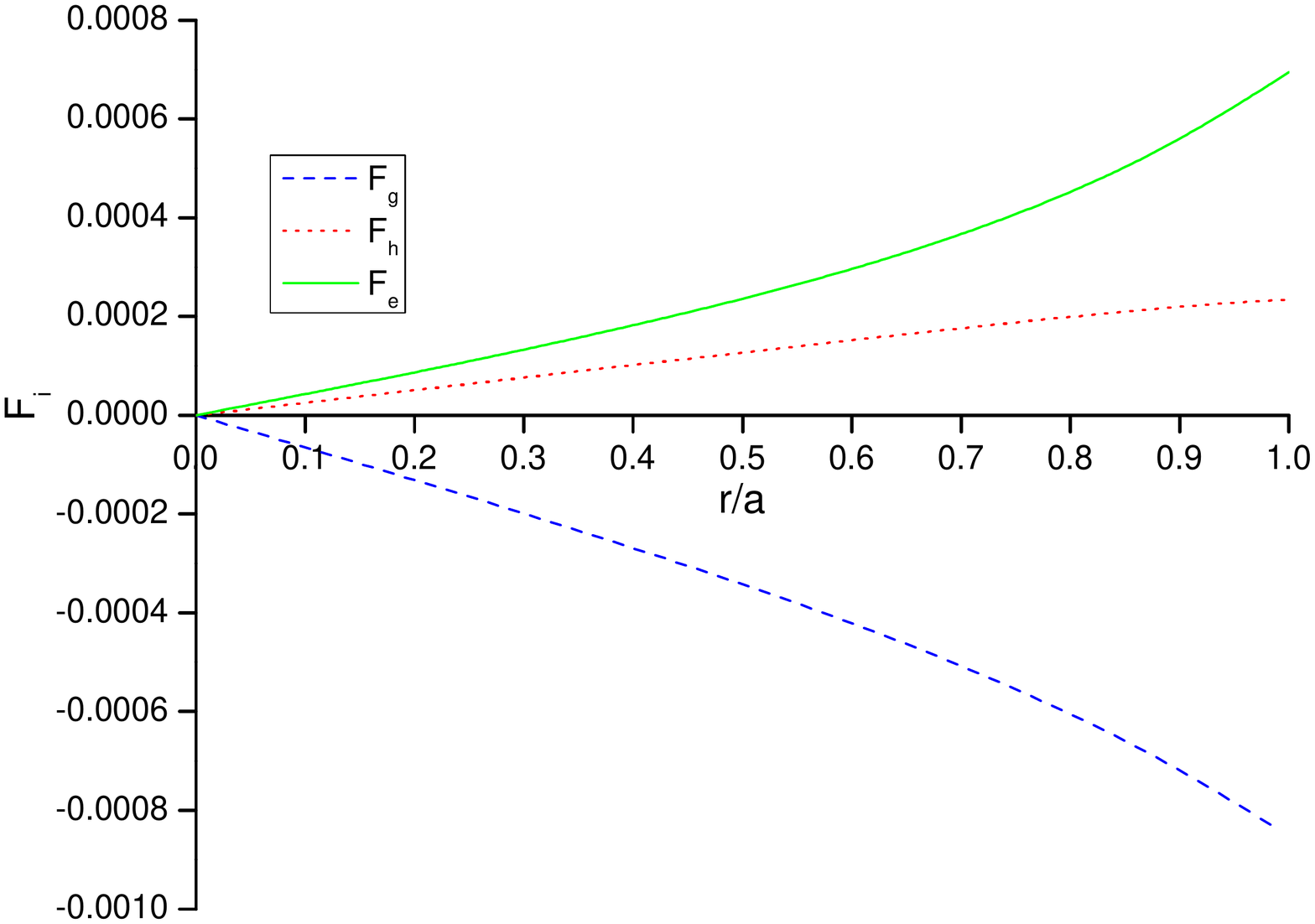}
\caption{Behavior of different forces(in $ km^{-3} $) vs. fractional radius r/a for the compact objects PSR J1614-2230,4U 1608-52, SAX J1808.4-3658, 4U 1538-52,SMC X-1,Her X-1 and Cen X-3. For this figure we have used the numerical values of physical parameters and constants are as follows: (i) K=0.0000135,b = 0.09, $C=-7.455\times10^{-8}km^{-2}$, $\eta^{2}$=6, M=$01.97M_{\odot}$ and $ a $ = 9.69 km for PSR J1614-2230(first row left),(ii) K = 0.0000209, b = 0.0903, $C=-1.1446\times10^{-7}km^{-2}$,$ \eta^{2} $=4.9, M =$01.74M_{\odot}$ , and $ a $ =9.3 km for 4U 1608-52(first row middle),(iii)K =0.0000319, b = 0.09,$C=-1.5664\times10^{-7}km^{-2}$,$ \eta^{2} $=3.9, M =$0.9M_{\odot}$ , and $ a $ =7.951 km for SAX J1808.4-3658(first row right),(iv)K = 0.0000296, b = 0.0903, $C=-1.4545\times10^{-7}km^{-2}$,$ \eta^{2} $=4, M =$0.87M_{\odot}$ , and $ a $ =7.866 km for 4U 1538-52(second row left),(v) K = 0.000023, b = 0.1,$C=-1.154\times10^{-7}km^{-2}$,$ \eta^{2} $=5, M =$1.29M_{\odot}$ , and $ a $ =8.831 km for SMC X-1(second row middle),(vi) K = 0.0000277, b = 0.06, $C=-1.2193\times10^{-7}km^{-2}$,$\eta^{2}$=2.9, M = $0.85M_{\odot}$ and $ a $ = 8.1 km for Her X-1(second row right),(vii) K = 0.000018, b = 0.09,$C=-9.4971\times10^{-8}km^{-2}$,$ \eta^{2} $=5, M =$1.49M_{\odot}$ , and $ a $ =9.178 km for Cen X-3(bottom)}\label{f2}
\end{center}
\end{figure}
\begin{eqnarray}
 -\frac{M_G(\rho+p)}{r^2}e^{\frac{\lambda-\nu}{2}}-\frac{dp}{dr}+
 \sigma \frac{q}{r^2}e^{\frac{\lambda}{2}} =0,\label{29} 
 \end{eqnarray}
where $M_G$ represents the gravitational mass and defined as:
\begin{eqnarray}
M_G(r)=\frac{1}{2}r^2 \nu^{\prime}e^{(\nu - \lambda)/2}.\label{30}
\end{eqnarray}
Substituting the value of $M_G(r)$ in equation (\ref{29}), we get
\begin{eqnarray}
-\frac{\nu'}{2}(\rho+p)-\frac{dp}{dr}+\sigma \frac{q}{r^2}e^{\frac{\lambda}{2}} =0,  \label{31}
\end{eqnarray}
The equation (\ref{31}) can be expressed into three unique segments, gravitational $(F_g)$, hydrostatic $(F_h)$ and electric $(F_e)$, which are defined as: 
\begin{eqnarray}
F_g=-\frac{\nu'}{2}(\rho+p)=\dfrac{Z'}{2\pi Z}(\rho+p)\label{32}
\end{eqnarray}
\begin{eqnarray}
F_h=-\frac{dp}{dr}=-\dfrac{1}{\pi}\left[{\dfrac{L}{(E2\times E5)^{2}}-\dfrac{2\sqrt{Cx}}{K(1+Y^{2})^{2}}+E7\times E8+E9\times E10}\right]  \label{33}
\end{eqnarray}
\begin{eqnarray}
F_e=\sigma \frac{q}{r^2}e^{\frac{\lambda}{2}}= \frac{1}{8\,\pi\,r^4}\,\frac{dq^2}{dr}=\dfrac{1}{2\pi}\left[E7\times E8+E9\times E10 ~\right]\label{34}\nonumber \\
\end{eqnarray}
We can see the behavior of the generalized
TOV equations by figure (\ref{f2}) and the system is counterbalanced by three different forces, e.g, gravitational force $(F_g)$, hydrostatic force$(F_h)$ and electric
force $(F_e)$. This conclude that the system attains a static equilibrium.
\section{Energy Conditions}
\begin{figure}[h]
\begin{center}
\includegraphics[width=5cm]{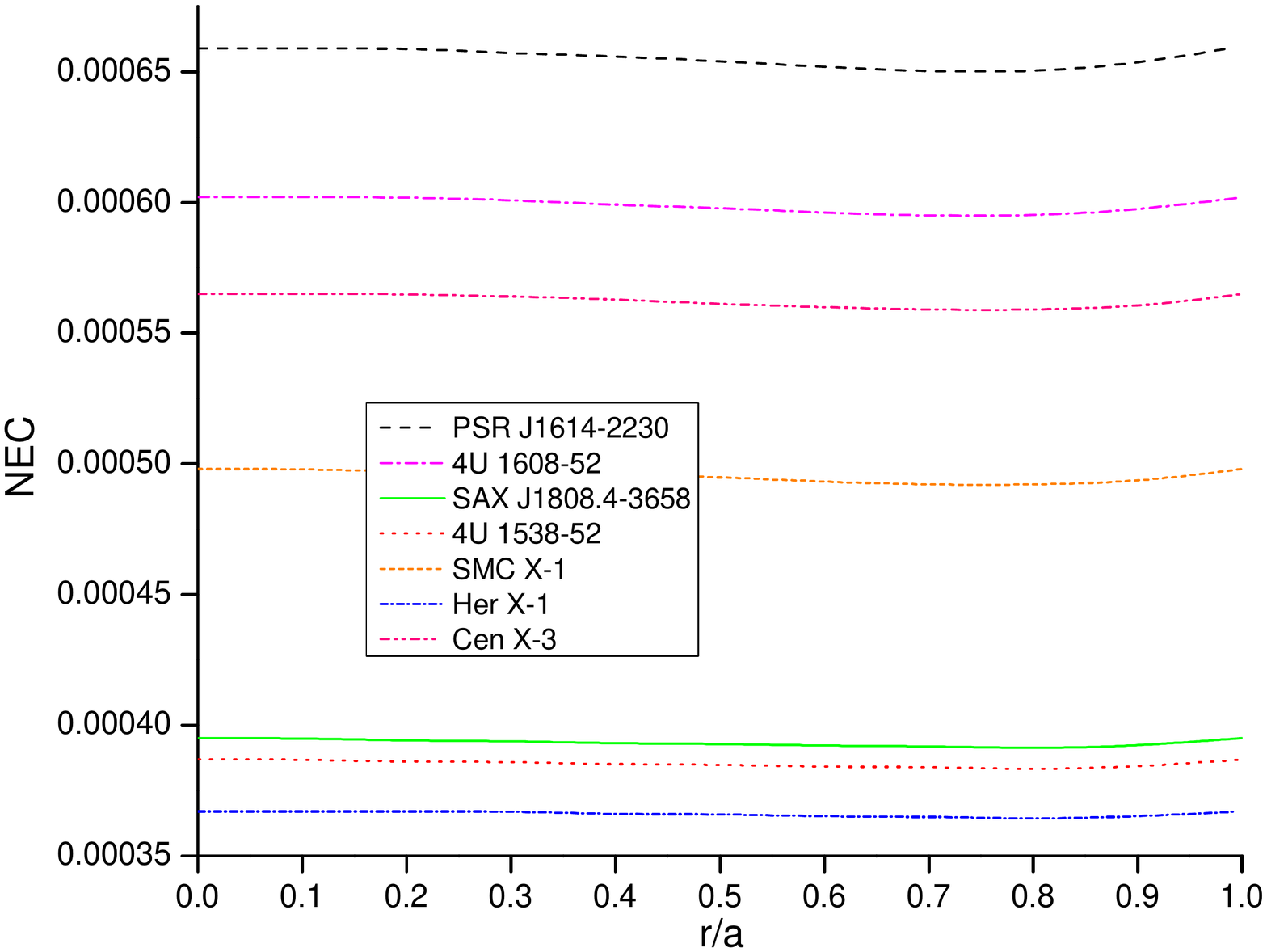}\includegraphics[width=5cm]{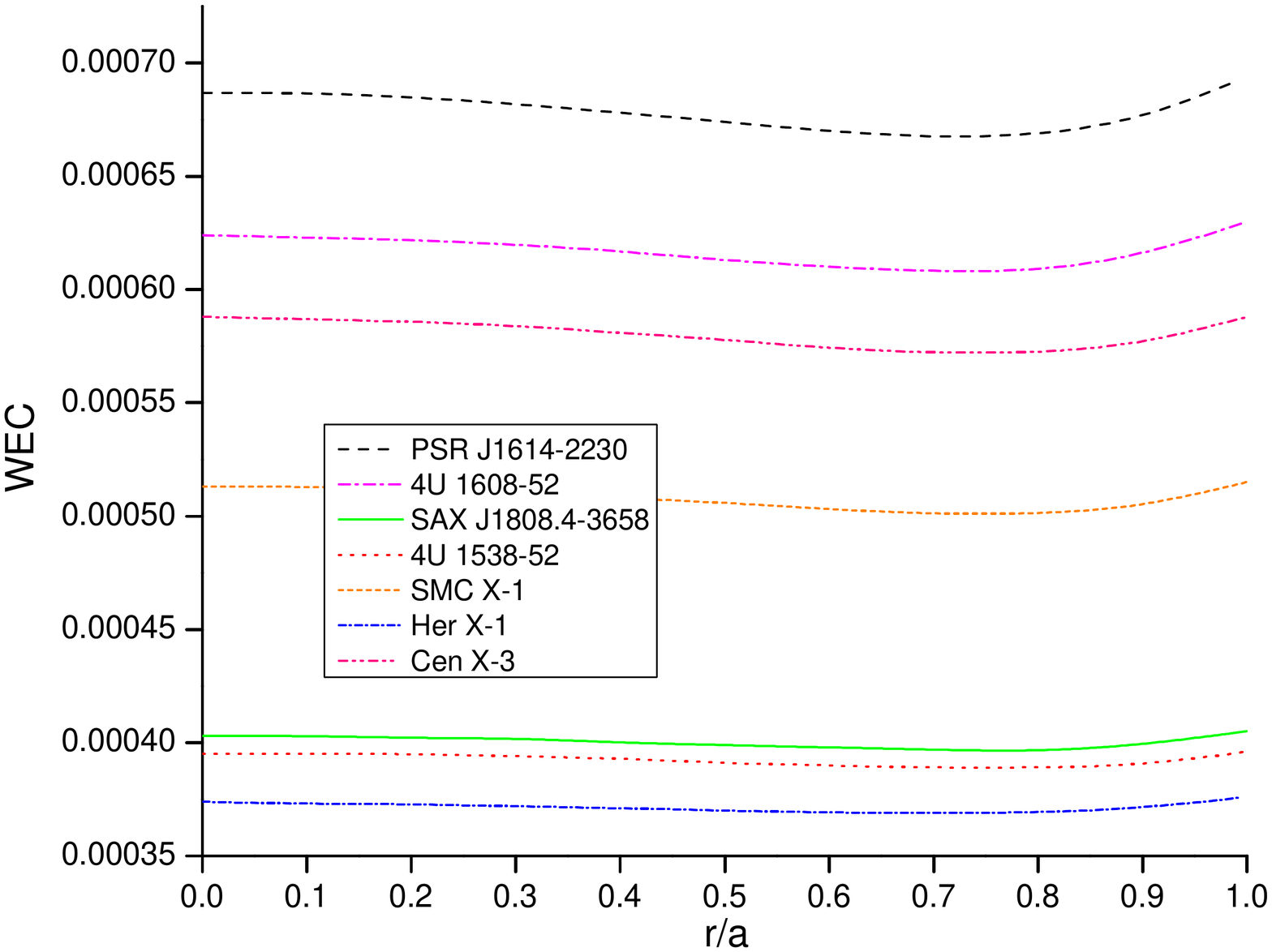}\includegraphics[width=5cm]{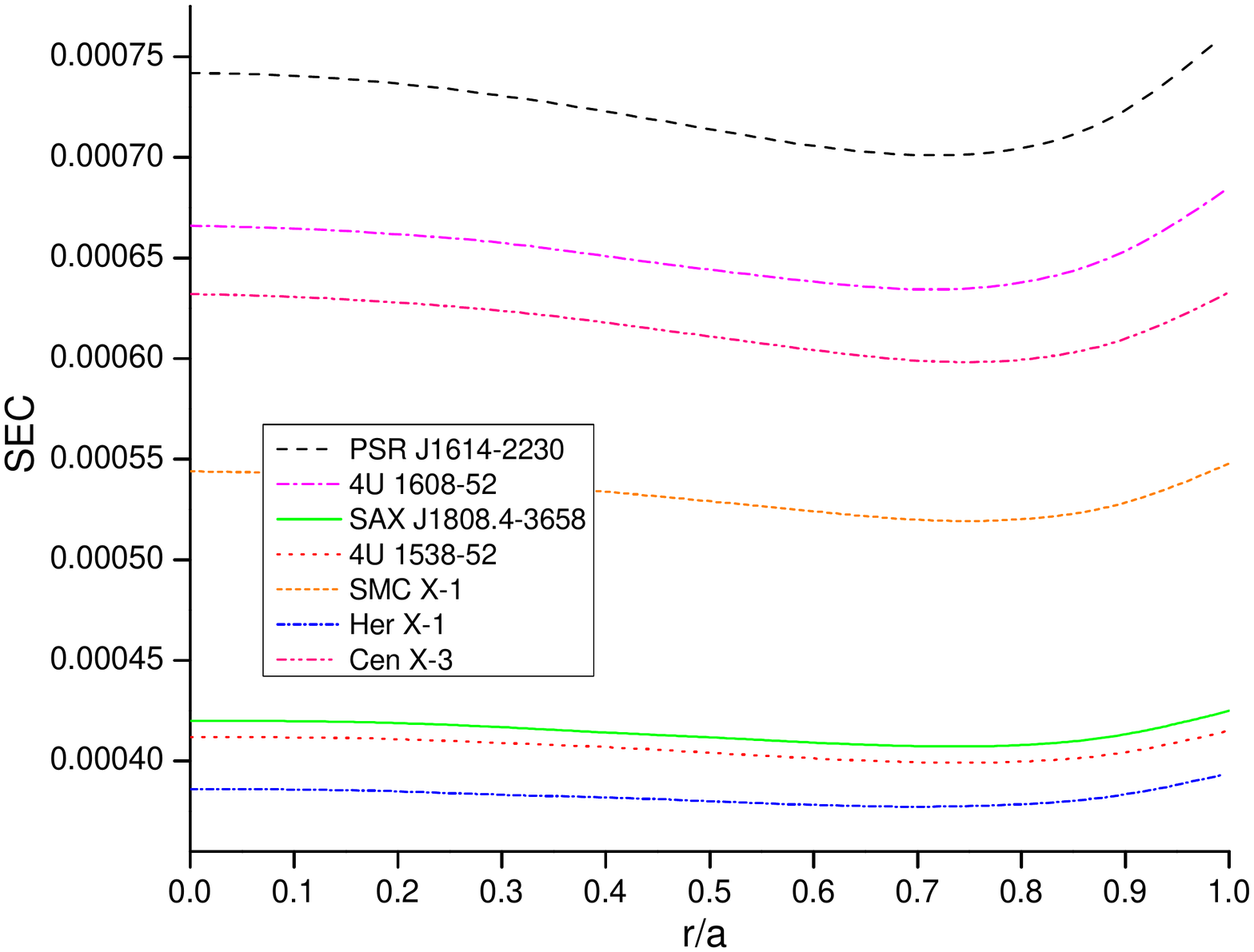}
\caption{Behavior of Energy conditions( in $ km^{-2} $) vs. fractional radius r/a for the compact stars PSR J1614-2230,4U 1608-52, SAX J1808.4-3658, 4U 1538-52,SMC X-1,Her X-1 and Cen X-3. In this figure we have used same data set values of physical parameters and constants which are the same in Fig.\ref{f1}} \label{f3}
\end{center}
\end{figure}

Here we analyze the energy conditions according to relativistic classical field theories of gravitation. In the context of GR the energy conditions are local inequalities that process a relation between matter density and pressure obeying certain restrictions.The charged fluid sphere should satisfy the three energy conditions (i) strong energy condition
(SEC),(ii) weak energy condition (WEC) and (iii)  null energy condition (NEC). For satisfying the above energy conditions, the
following inequalities must hold simultaneously inside the charged fluid sphere: \\
Null energy condition (NEC): $\rho+\frac{q^2}{r^4\,\pi} \geq 0 $\\
Weak energy condition (WEC ):$ \rho+p+\frac{q^2}{r^4\,4\pi} \geq 0$\\
Strong energy condition (SEC):$ \rho+3 p +\frac{q^2}{r^4\,2\pi} \geq 0.$\\
The nature of energy conditions for the specific stellar configuration as shown in Fig.\ref{f3}, that are satisfied for our proposed model.
\vspace{2mm}
\begin{table}
\caption{\label{ta1}Values of different physical parameters of PSR J1614-2230}
\begin{tabular}{ccccccc}
\toprule
\multicolumn{7}{c}{$  K=0.0000135,~~ C=-7.455\times10^{-8}km^{-2},~~\eta^{2}=6,~~b=0.09,~~Zo=0.807893,~~Za=0.513176$} \\
\hline
r/a & p$ (km^{-2}) $ & $\rho (km^{-2}) $ & q$ (km) $ & $dp/c^{2}d\rho$ &	$ p/\rho $ & $\gamma$ \\ 
\hline
0&	2.7585$ \times 10^{-5} $&	0.000659&	0&	0.583119&	0.041827&	14.524215\\
0.1&	2.7255$ \times 10^{-5} $&	0.000659&	0.004581&	0.578524&	0.041342&	14.572066\\
0.2&	2.6272$ \times 10^{-5} $&	0.000659&	0.03682&	0.564766&	0.039891&	14.722468\\
0.3&	2.4645$ \times 10^{-5} $&	0.000657&	0.125268&	0.541926&	0.037488&	14.997957\\
0.4&	2.2397$ \times 10^{-5} $&	0.000656&	0.300384&	0.510103&	0.034157&	15.443954\\
0.5&	1.9563$ \times 10^{-5} $&	0.000653&	0.595859&	0.46936&	0.029939&	16.146318\\
0.6&	1.6191$ \times 10^{-5} $&	0.00065&	1.050501&	0.419633&	0.024895&	17.275441 \\
0.7&	1.2362$ \times 10^{-5} $&	0.000646&	1.711198&	0.360575&	0.019124&	19.215468\\
0.8&	8.2021$ \times 10^{-6} $&	0.000641&	2.63799&	0.291288&	0.01279&	23.065459\\
0.9&	3.9353$ \times 10^{-6} $&	0.000634&	3.913638&	0.209843&	0.006202&	34.042794\\
1&	0&	0.000625&	5.663978&	0.11233&	0&	Inf\\
\bottomrule
\end{tabular}
\end{table}
\vspace{1mm}
\begin{table}
\caption{Value of different physical parameter of 4U 1608-52}
\label{ta2}
\begin{tabular}{ccccccc}
\toprule
\multicolumn{7}{c}{$  K=0.0000209, C=-1.1446\times10^{-7}km^{-2}, \eta^{2}=4.9, b=0.0903, Zo=0.691097,Za=0.0.445919 $} \\
\hline
r/a & p$ (km^{-2}) $ & $\rho (km^{-2}) $ & q$ (km) $ & $dp/c^{2}d\rho$ &	$ p/\rho $ & $\gamma$ \\ 
\hline
0&	$ 2.2941\times 10^{-5} $&	0.000654&	0&	0.580658&	0.035075&	17.135305\\
0.1&	$ 2.2671\times 10^{-5} $&	0.000654&	0.004033&	0.577187&	0.034673&	17.223585\\
0.2&	$ 2.1864\times 10^{-5} $&	0.000653&	0.032405&	0.566776&	0.033471&	17.499985\\
0.3&	$ 2.053\times 10^{-5} $&	0.000652&	0.110168&	0.549433&	0.03148&	18.002797\\
0.4&	$ 1.8686\times 10^{-5} $&	0.000651&	0.2639&	   0.525142&	0.02872&	18.81021\\
0.5&	$ 1.6357\times 10^{-5} $&	0.000649&	0.522728&	0.493821&	0.025222&	20.072895\\
0.6&	$ 1.3584\times 10^{-5} $&	0.000646&	0.919748&	0.45525&	0.021035&	22.097382\\
0.7&	$ 1.0428\times 10^{-5} $&	0.000642&	1.494186&	0.408948&	0.016236&	25.595997\\
0.8&	$ 6.9842\times 10^{-6} $&	0.000638&	2.294923&	0.353977&	0.01095&	32.680025\\
0.9&	$ 3.4102\times 10^{-6} $&	0.000632&	3.386797&	0.28859&   0.005396&	53.766753\\
1&	0&	0.000624&	4.863006&	0.209584&	0&	Inf
\\ \bottomrule
\end{tabular}
\end{table}
\newpage
\begin{table}
\caption{Value of different physical parameter of SAX J1808.4-3658}
\label{ta3}
\begin{tabular}{ccccccc}
\toprule
\multicolumn{7}{c}{$  K=0.0000319, C=-1.5664\times10^{-7}km^{-2}, \eta^{2}=3.9, b=0.09, Zo=0.390495,Za=0.268166$} \\
\hline
r/a & p$ (km^{-2}) $ & $\rho (km^{-2}) $ & q$ (km) $ & $dp/c^{2}d\rho$ &	$ p/\rho $ & $\gamma$ \\ 
\hline
0&	$ 1.2505\times 10^{-5} $&	0.000586&	0&	0.669955&	0.02133&	32.079514\\
0.1&	$ 1.2368\times 10^{-5} $&	0.000586&	0.002149&	0.668107&	0.021101&	32.330121\\
0.2&	$ 1.196\times 10^{-5} $&	0.000586&	0.017239&	0.662554&	0.020417&	33.113282\\
0.3&	$ 1.1283\times 10^{-5} $&	0.000585&	0.058458&	0.653264&	0.019282&	34.533258\\
0.4&	$ 1.0343\times 10^{-5} $&	0.000584&	0.139505&	0.640178&	0.017702&	36.804817\\
0.5&	$ 9.1483\times 10^{-6} $&	0.000583&	0.274902&	0.623194&	0.015688&	40.346788\\
0.6&	$ 7.7107\times 10^{-6} $&	0.000582&	0.480386&	0.60215&	0.013257&	46.023518\\
0.7&	$ 6.0477\times 10^{-6} $&	0.00058&	0.773419&	0.576791&	0.010431&	55.874965\\
0.8&	$ 4.1825\times 10^{-6} $&	0.000578&	1.173914&	0.546727&	0.007242&	76.041701\\
0.9&	$ 2.1498\times 10^{-6} $&	0.000575&	1.705293&	0.511364&	0.00374&	137.243765\\
1&	0&	0.000571&	2.39613&	0.469797&	0&	Inf
\\ \bottomrule
\end{tabular}
\end{table}
\vspace{1mm}
\begin{table}
\caption{Value of different physical parameter of 4U 1538-52}
\label{ta4}
\begin{tabular}{ccccccc}
\toprule
\multicolumn{7}{c}{$  K=0.0000296, C=-1.4545\times10^{-7}km^{-2}, \eta^{2}=4, b=0.0903, Zo=0.384478,Za=0.265254$} \\
\hline
r/a & p$ (km^{-2}) $ & $\rho (km^{-2}) $ & q$ (km) $ & $dp/c^{2}d\rho$ &	$ p/\rho $ & $\gamma$  \\ 
\hline
0&	$ 1.2854\times 10^{-5} $&	0.000587&	0&	0.729746&	0.021904&	34.045791\\
0.1&	$ 1.2715\times 10^{-5} $&	0.000587&	0.002072&	0.727689&	0.021671&	34.306902\\
0.2&	$ 1.2298\times 10^{-5} $&	0.000586&	0.016621&	0.72151&	0.020973&	35.12295\\
0.3&	$ 1.1607\times 10^{-5} $&	0.000586&	0.056356&	0.711185&	0.019815&	36.602787\\
0.4&	$ 1.0647\times 10^{-5} $&	0.000585&	0.134467&	0.696664&	0.018202&	38.970655\\
0.5&	$ 9.4263\times 10^{-6} $&	0.000584&	0.264921&	0.677864&	0.016145&	42.663925\\
0.6&	$ 7.9548\times 10^{-6} $&	0.000582&	0.462818&	0.654646&	0.013658&	48.585353\\
0.7&	$ 6.2492\times 10^{-6} $&	0.000581&	0.744878&	0.626788&	0.010762&	58.865847\\
0.8&	$ 4.3313\times 10^{-5} $&	0.000578&	1.130102&	0.593945&	0.007487&	79.920526\\
0.9&	$ 2.2325\times 10^{-5} $&	0.000576&	1.640744&	0.555585&	0.003877&	143.84548\\
1&	0&	0.000573&	2.303804&	0.510896&	0&	Inf
\\ \bottomrule
\end{tabular}
\end{table}
\vspace{1mm}
\begin{table}
\caption{Value of different physical parameter of SMC X-1}
\label{ta5}
\begin{tabular}{ccccccc}
\toprule
\multicolumn{7}{c}{$  K=0.000023, C=-1.154\times10^{-7}km^{-2}, \eta^{2}=5, b=0.1, Zo=0.53033,Za=0.353253$} \\
\hline
r/a & p$ (km^{-2}) $ & $\rho (km^{-2}) $ & q$ (km) $ & $dp/c^{2}d\rho$ &	$ p/\rho $ & $\gamma$ \\ 
\hline
0&  	$ 1.8593\times 10^{-5} $&	0.000599&	0&	0.674146&	0.03103&	22.399988\\
0.1&	$ 1.8387\times 10^{-5} $&	0.000599&	0.003005&	0.670942&	0.030692&	22.531393\\
0.2&	$ 1.7767\times 10^{-5} $&	0.000599&	0.024126&	0.661333&	0.029681&	22.942466\\
0.3&	$ 1.6741\times 10^{-5} $&	0.000598&	0.081913&	0.645324&	0.028004&	23.689209\\
0.4&	$ 1.5319\times 10^{-5} $&	0.000597&	0.195834&	0.622911&	0.025673&	24.886618\\
0.5&	$ 1.3514\times 10^{-5} $&	0.000595&	0.38686&	0.594054&	0.022705&	26.758037\\
0.6&	$ 1.1348\times 10^{-5} $&	0.000593&	0.678235&	0.558639&	0.019129&	29.76226\\
0.7&	$ 8.8539\times 10^{-5} $&	0.000591&	1.096575&	0.516417&	0.014986&	34.977079\\
0.8&	$ 6.0759\times 10^{-5} $&	0.000588&	1.673515&	0.466911&	0.010338&	45.633544\\
0.9&	$ 3.0857\times 10^{-5} $&	0.000584&	2.448385&	0.409257&	0.005285&	77.846602\\
1&	0	&0.000579&	3.472844&	0.341948&	0&	Inf
\\ \bottomrule
\end{tabular}
\end{table}
\vspace{1mm}
\begin{table}
\caption{Value of different physical parameter of Her X-1}
\label{ta6}
\begin{tabular}{ccccccc}
\toprule
\multicolumn{7}{c}{$  K=0.0000277, C=-1.2193\times10^{-7}km^{-2}, \eta^{2}=2.9, b=0.06, Zo=0.351093,Za=0.242073$} \\
\hline
r/a & p$ (km^{-2}) $ & $\rho (km^{-2}) $ & q$ (km) $ & $dp/c^{2}d\rho$ &	$ p/\rho $ & $\gamma$ \\ 
\hline
0&	$ 8.8592\times 10^{-6} $&	0.000526&	0&	0.785045&	0.016853&	47.367521\\
0.1&	$ 8.7621\times 10^{-6} $& 	0.000526&	0.00196&	0.783757&	0.016671&	47.798122\\
0.2&	$ 8.4702\times 10^{-6} $&	0.000525&	0.015725&	0.779875&	0.016125&	49.14376\\
0.3&	$ 7.9867\times 10^{-6} $&	0.000525&	0.053307&	0.773342&	0.01522&	51.583661\\
0.4&	$ 7.3162\times 10^{-6} $&	0.000524&	0.127154&	0.764058&	0.013962&	55.487418\\
0.5&	$ 6.4649\times 10^{-6} $&	0.000523&	0.250411&	0.751865&	0.012361&	61.576861\\
0.6&	$ 5.4425\times 10^{-6} $&	0.000522&	0.437239&	0.736529&	0.010431&	71.344062\\
0.7&	$ 4.2618\times 10^{-6} $&	0.00052&	0.703237&	0.717719&	0.008193&	88.315903\\
0.8&	$ 2.9421\times 10^{-6} $&	0.000518&	1.066009&	0.694956&	0.005676&	123.122389\\
0.9&	$ 1.5084\times 10^{-6} $&	0.000516&	1.545994&	0.667566&	0.002924&	228.989058\\
1&	0&	0.000513&	2.167694&	0.634581&	0&	Inf
\\ \bottomrule
\end{tabular}
\end{table}\vspace{1mm}
\begin{table}
\caption{Value of different physical parameter of Cen X-3}
\label{ta7}
\begin{tabular}{ccccccc}
\toprule
\multicolumn{7}{c}{$  K=0.000018, C=-9.4971\times10^{-8}km^{-2}, \eta^{2}=5, b=0.09, Zo=0.64214,Za=0.419462$} \\
\hline
r/a & p$ (km^{-2}) $ & $\rho (km^{-2}) $ & q$ (km) $ & $dp/c^{2}d\rho$ &	$ p/\rho $ & $\gamma$ \\ 
\hline
0&	$ 2.4771\times 10^{-5} $&	0.00063&	0&	0.895387&	0.039312&	23.672047\\
0.1&	$ 2.4491\times 10^{-5} $&	0.00063&	0.003602&	0.890504&	0.03888&	23.794584\\
0.2&	$ 2.3657\times 10^{-5} $&	0.000629&	0.028928&	0.875877&	0.037587&	24.178474\\
0.3&	$ 2.2275\times 10^{-5} $&	0.000628&	0.098299&	0.851556&	0.035443&	24.877758\\
0.4&	$ 2.0359\times 10^{-5} $&	0.000627&	0.235294&	0.817605&	0.032463&	26.003314\\
0.5&	$ 1.7931\times 10^{-5} $&	0.000625&	0.465583&	0.774064&	0.028673&	27.770227\\
0.6&	$ 1.5021\times 10^{-5} $&	0.000623&	0.818063&	0.720884&	0.024111&	30.619701\\
0.7&	$ 1.1677\times 10^{-5} $&	0.00062&	1.326537&	0.657833&	0.018834&	35.586676\\
0.8&	$ 7.9693\times 10^{-6} $&	0.000616&	2.032382&	0.584338&	0.012932&	45.769141\\
0.9&	$ 4.0109\times 10^{-6} $&	0.000611&	2.989146&	0.499213&	0.00656&	76.597939\\
1&	0&	0.000605&	4.271195&	0.400191&	0&	Inf
\\ \bottomrule
\end{tabular}
\end{table}
\vspace{1mm}
\begin{table}
\caption{Numerical values of radius($ a $) $ M(M_{\odot}),$ central density, surface density,central pressure and mass-radius ratio of compact star candidates.}
\label{ta8}
\begin{tabular}{lllllll} 
\toprule
Compact star & a($km$) & M($M_{\odot})$ & Central density & Surface density & Central pressure &   M/a \\
& & & ($g/cm^{3}  $)& ($g/cm^{3} $)& ($ dyne/cm^{2} $)& \\
\hline 
  PSR J1614-2230 & 9.69 & 1.97 & 8.886$ \times 10^{14} $ & 8.4233$ \times 10^{14} $ & 3.3453$ \times 10^{34} $ & 0.16314 \\ 
 4U 1608-52 & 9.3 & 1.74 & 8.8133$ \times 10^{14} $ & 8.41008$ \times 10^{14} $ & 2.7821$ \times 10^{34} $ &0.2694 \\ 
 SAX J1808.4-3658 &  7.95 & 0.9 & 7.8997$ \times 10^{14} $ & 7.6997$ \times 10^{14} $ & 1.5165$ \times 10^{34} $ & 0.16696\\
 4U 1538-52 & 7.866 & 0.87 & 7.9078$ \times 10^{14} $ & 7.7151$ \times 10^{14} $ & 1.5588$ \times 10^{34} $ &0.1631\\
 SMC X-4 & 8.831 & 1.29 & 8.0744$ \times 10^{14} $ & 7.8018$ \times 10^{14} $ & 2.2549$ \times 10^{34} $ & 0.2154 \\ 
 Her X-1 & 8.1 & 0.85 & 7.0836$ \times 10^{14} $ & 6.9144$ \times 10^{14} $ & 1.0744$ \times 10^{34} $ &0.1547 \\ 
Cen X-3 & 9.165 & 1.49 & 8.4906$ \times 10^{14} $ & 8.1524$ \times 10^{14} $ & 3.004$ \times 10^{34} $ & 0.2487 \\  \bottomrule
\end{tabular}
\end{table}
\begin{figure}[h]
\begin{center}
\includegraphics[width=6cm]{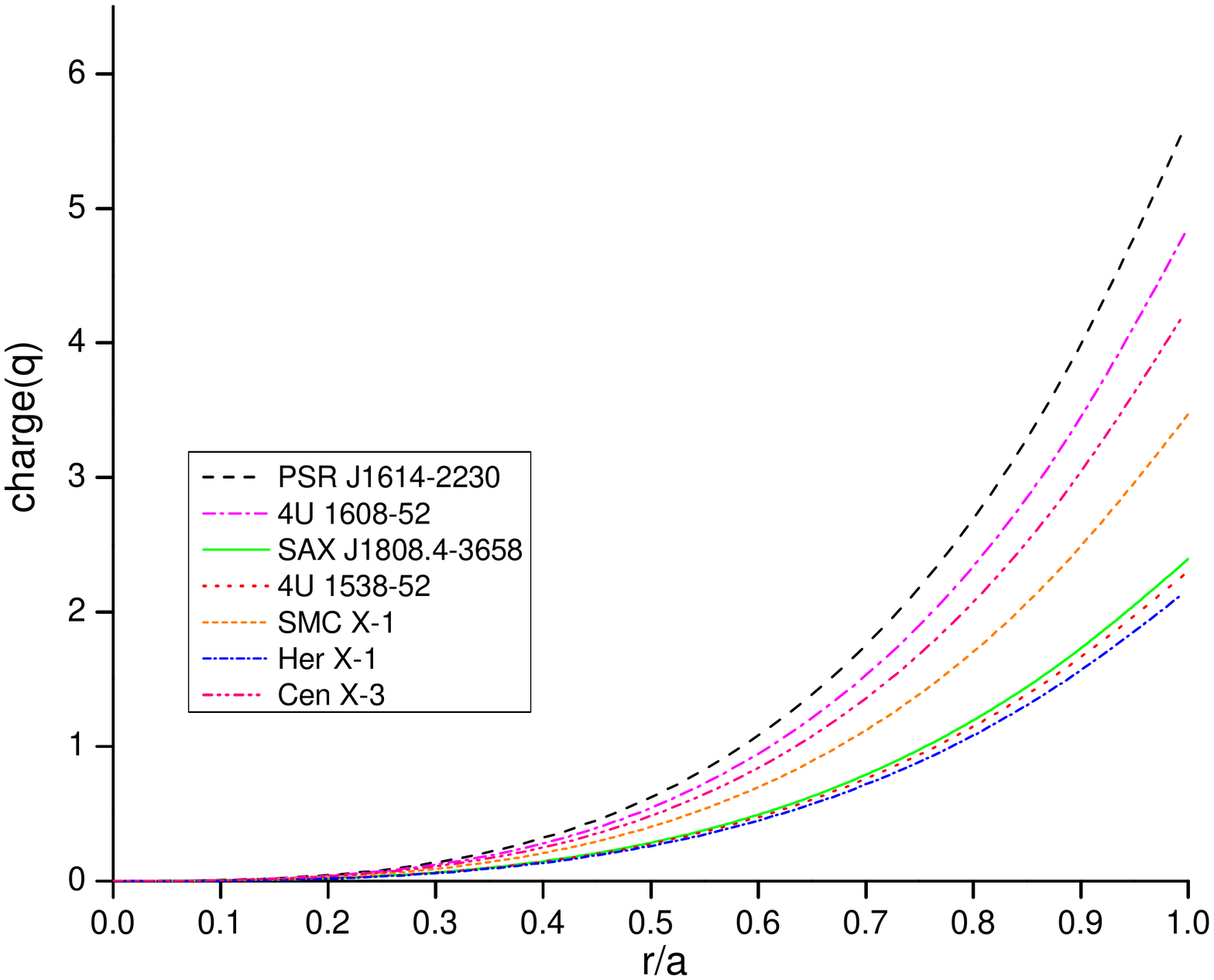}
\caption{Behavior of Charge(q in $ km $) vs. fractional radius
r/a for the compact stars PSR J1614-2230,4U 1608-52, SAX J1808.4-3658, 4U 1538-52,SMC X-1,Her X-1 and Cen X-3.In this figure we have used same data set values of physical parameters and constants which are the same in Fig.\ref{f1}}\label{f4}
\end{center}
\end{figure}
\section{Electric Charge}
Varela et al.\cite{varela} have shown that fluid spheres with net charge contain fluid elements with unbounded proper charge density located at the fluid-vacuum interface and net charge can be huge$(10^{19}C)$. Ray et al.\cite{sray} have analyzed the impact of charge in compact stars considering the limit of the most extreme measure of the charge. They have demonstrated the global balance of the forces allow a huge charge$(10^{20}C)$ to be available in compact star.\par In this model we have found that the maximum  charge on the boundary is $ 6.605\times 10^{20} C $ and at the center is zero. We have plot the Fig.\ref{f4} for the charge $ q $ in the relativistic units(km).For coulombs unit, one has multiply these value by $ 1.1659\times 10^{20} C $.Thus in this model the net amount of charge is effective to balance the mechanism of the force.
\section{Surface Redshift}
The gravitational redshift $ Z_{s} $ within a static line element can be obtained as 
\begin{eqnarray}
Z_{s}=\sqrt{g_{tt}(a)}-1=\sqrt{1-\frac{2M}{a}+\frac{q^2}{a^2}}-1\label{35}
\end{eqnarray}
where $ g_{tt}(a)=e^{\nu(a)}=1-\frac{2M}{a}+\frac{q^2}{a^2} $\\
\begin{figure}[h]
\begin{center}
\includegraphics[width=6cm]{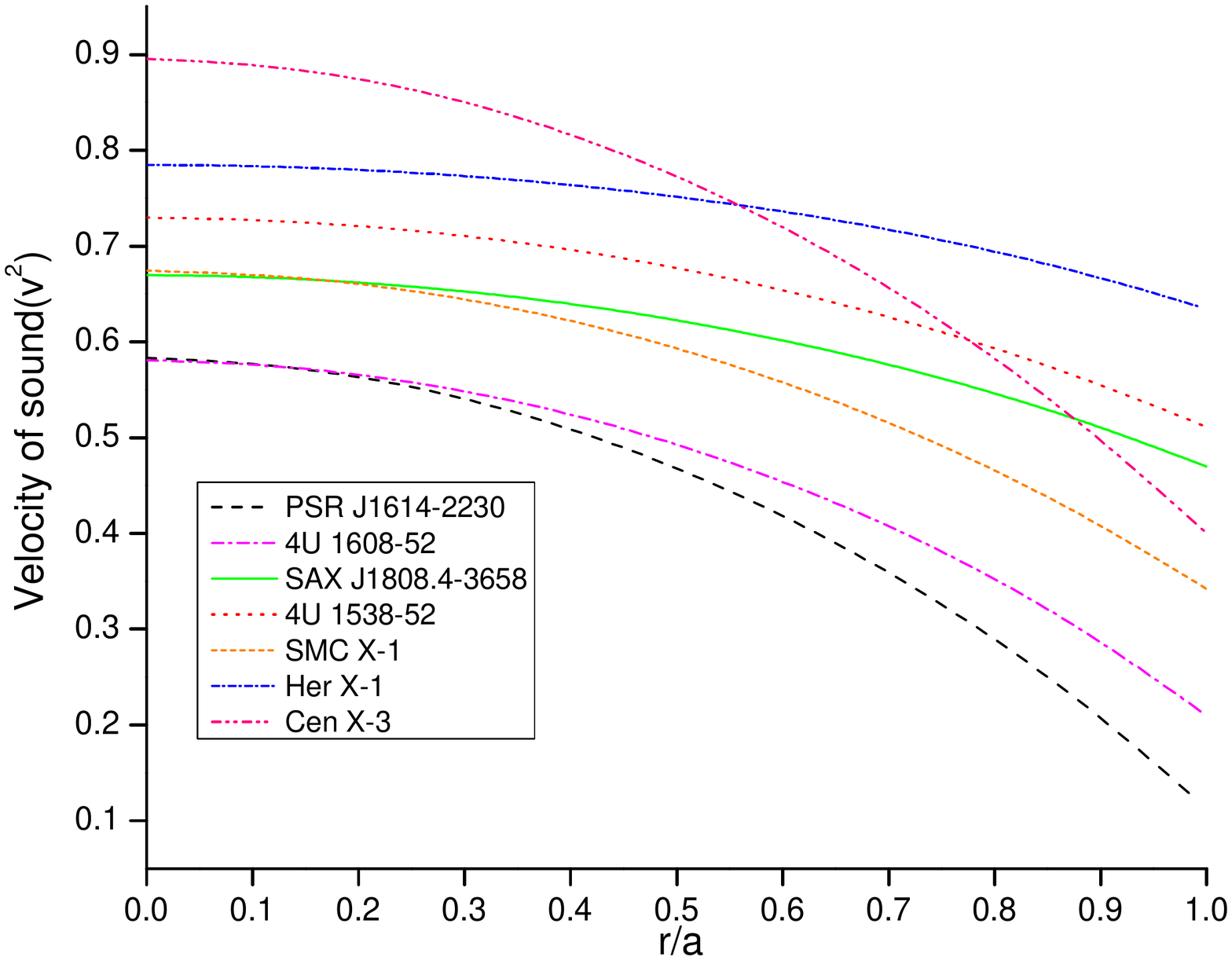}\includegraphics[width=6cm]{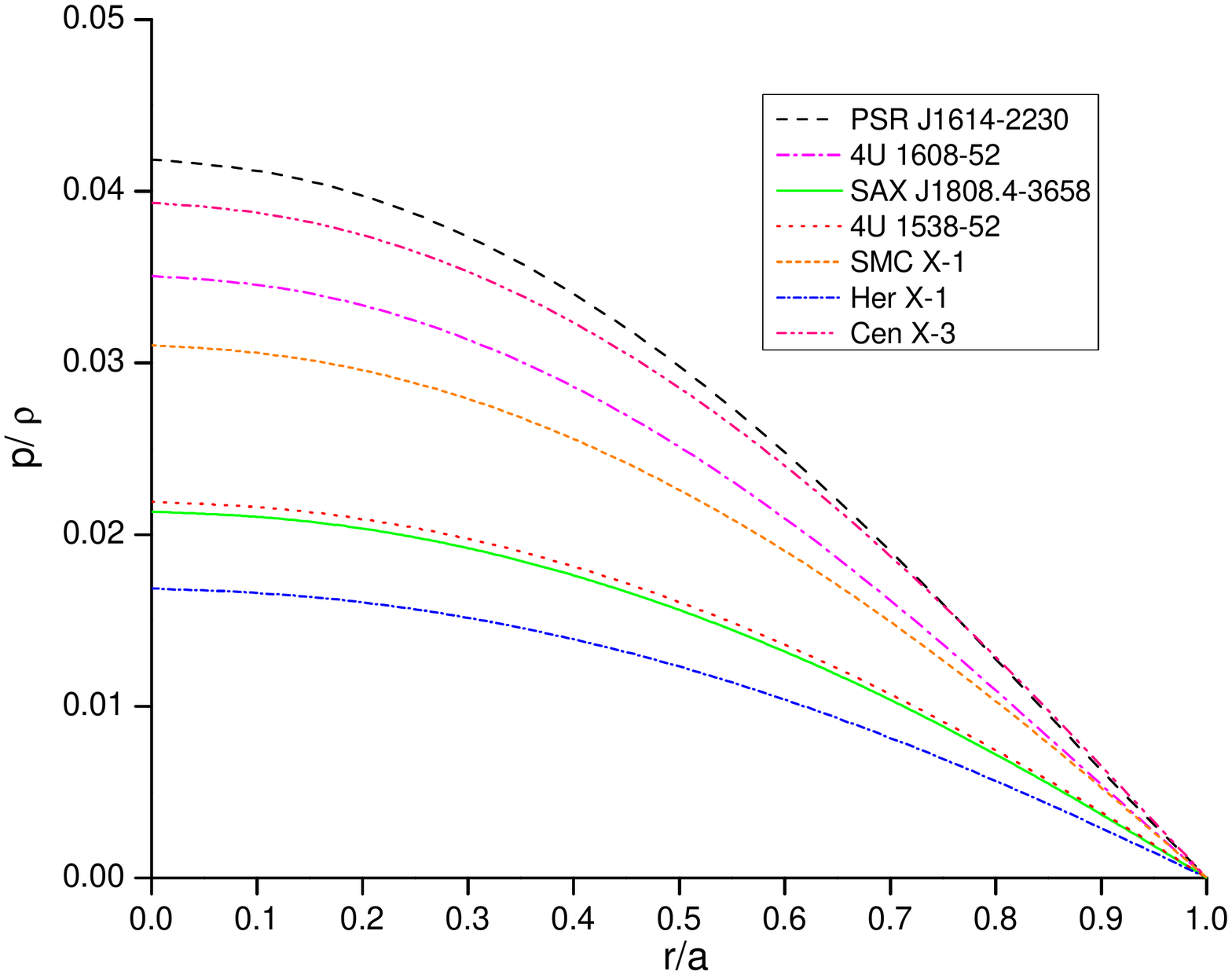}\\\includegraphics[width=6cm]{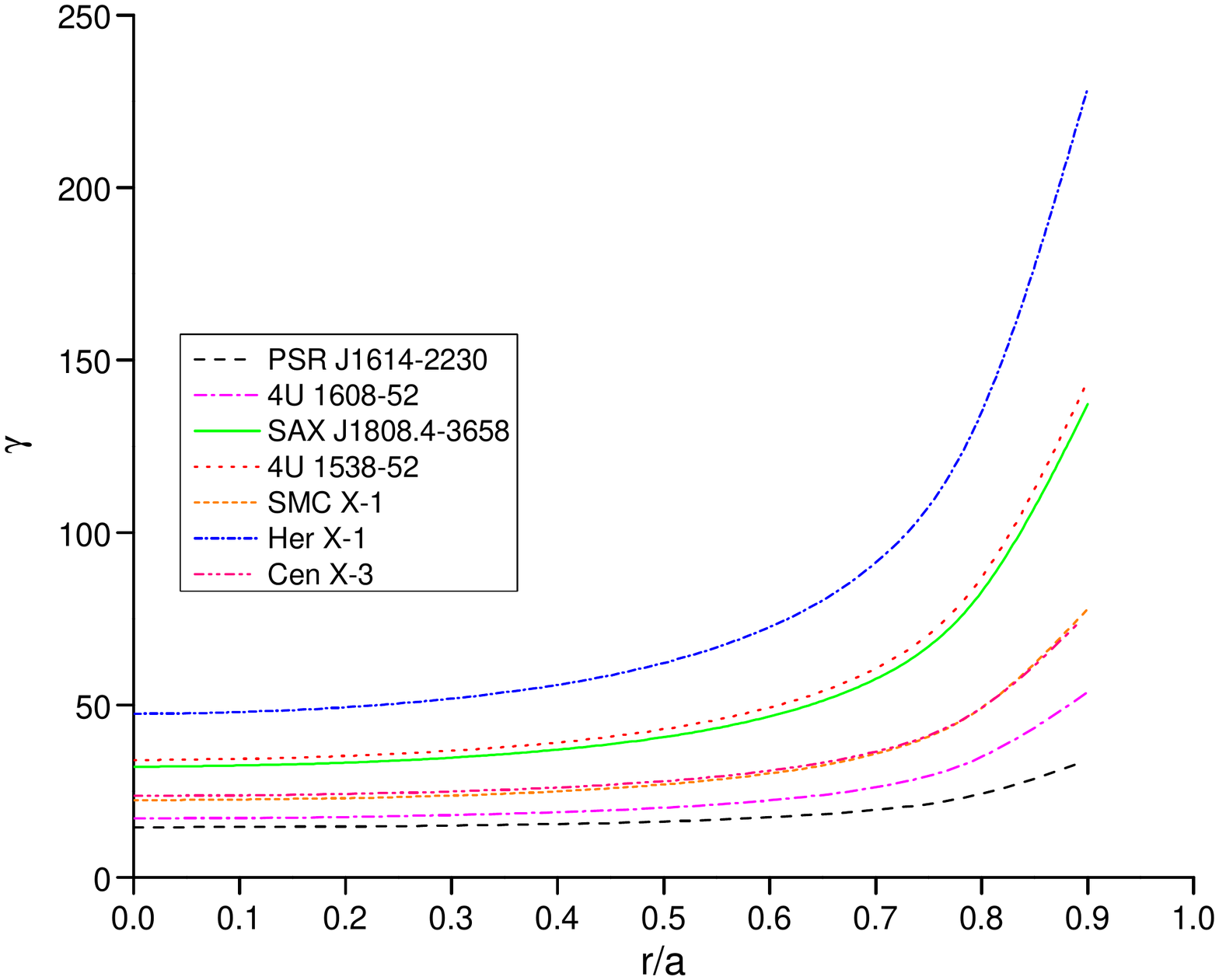}\includegraphics[width=6cm]{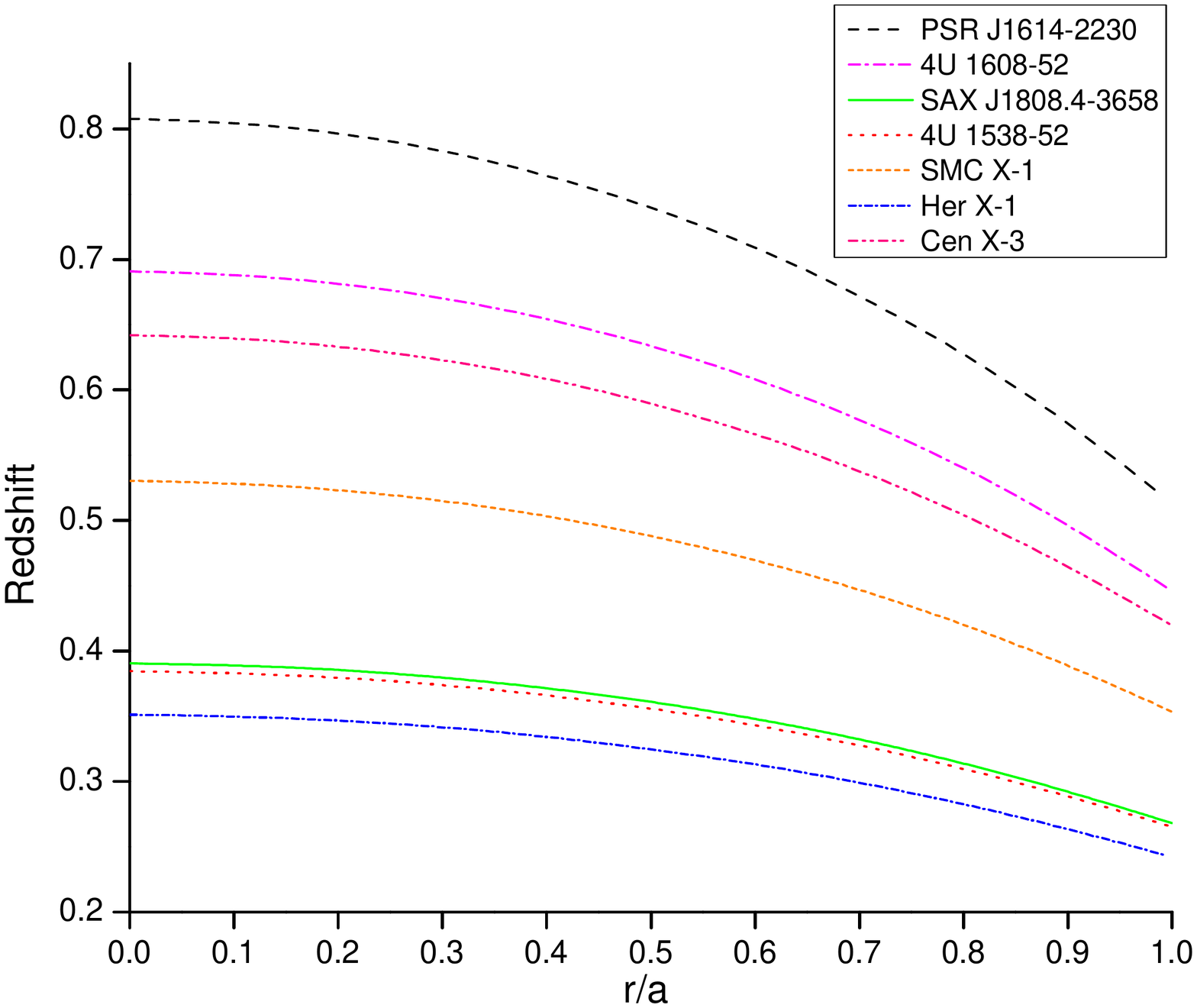}
\caption{Behaviour of Velocity of sound,density-pressure ratio,adiabatic constant and redshift vs. fractional radius
r/a for the compact stars PSR J1614-2230,4U 1608-52, SAX J1808.4-3658, 4U 1538-52,SMC X-1,Her X-1 and Cen X-3. For this figure we have used same data set values of physical parameters and constants which are the same in Fig.\ref{f1}}\label{f5}
\end{center}
\end{figure}
The maximum possible value of redshift should be at the center of the star and decrease with the increase of radius.Buchdahl\cite{buchdahl} and Straumann\cite{straumann} have shown that for an isotropic star the surface redshift $ Z_{s}\leq 2 $.For an anisotropic star Bohmer and Harko \cite{bohmer} showed that the surface redshift could be increased up to $ Z_{s}\leq 5 $. Ivanov\cite{Ivanov} modified the maximum value of redshift and showed that it could be as high as $ Z_{s}= 5.211 $. In this model we have  $ Z_{s}\leq 1 $ for compact stars PSR J1614-2230,4U 1608-52,SAX J1808.4-3658,4U 1538-52,SMC X-1, Her X-1 and Cen X-3.Also it is decreasing towards the boundary(see Fig.\ref{f5} bottom right).
\section{Causality and Well behaved condition}
Inside the fluid sphere the velocity of sound is less than the light, i.e. $ 0\leq v^{2} \leq \dfrac{dp}{d\rho}<1 $. According to Canuto\cite{canuto}, for well behaved nature of the charge solution, the velocity of sound should be monotonically decreasing towards the boundary with an ultra-high distribution of matter.From Fig.\ref{f5}(top right) it is verified that velocity of sound should monotonically decreasing.This imply our model for charge compact star is well behaved.
\section{Adiabatic Index}
The stability of relativistic isotropic fluid sphere depends on the adiabatic index $ \gamma $. Heintzmann and Hillebrandt\cite{heint} proposed that isotropic compact star models are stable if $ \gamma > 4/3 $ thought the star.In present model the adiabatic index defined by 
\begin{eqnarray}
\gamma = \big(\dfrac{p+\rho}{p} \big)\dfrac{dp}{d\rho}\label{36}
\end{eqnarray}
From Fig.\ref{f5}(bottom left)  we have seen that the value of $ \gamma $ is greater than $4/3$ and hence the model is stable.
\section{Harrison-Zeldovich-Novikov Stability Criterion}
The Harrison-Zeldovich-Novikov\cite{harison,novi} criterion states that the compact star to be stable if the its mass is increasing with increasing central density,i.e. $ \frac{dM}{d\rho_{0}}>0 $, however it is unstable if $ \frac{dM}{d\rho_{0}}<0 $. For this purpose we calculate the density of star at the center ($ \rho_{0}) $, mass $( M(a))$ and its gradient in terms of central density as \begin{eqnarray}
\rho_{0} =\dfrac{3C(K-1)}{8\pi K}\label{37}
\end{eqnarray}
\begin{eqnarray}
M(a) &=&\dfrac{4\pi \rho_{0}\,a^3 }{M_1(a)}\bigg[K-1+\dfrac{\pi \rho_{0}\,K\,a^2}{M_1(a)}\,M_2(a)\bigg]\label{38}\\
\dfrac{dM(a)}{d\rho_{0}}&=&\dfrac{12\pi\,a^3\,(K-1)}{(M_1(a))^2}\bigg[K-1+\dfrac{2\pi\,\rho_{0}K\,a^2}{M_1(a)}\,M_2(a)+\dfrac{4\pi\,(\rho_{0}\,K\,a)^2}{3(K-1)}\bigg(\dfrac{-30\pi\,a^2(1-K)^2}{M_1(a)}-M_3(a)\bigg)\bigg]\label{39}
\end{eqnarray}
where $ M_1(a)=8\pi\,\rho_{0}\,K\,a^2\,+3(K-1)\\
M_2(a)= \bigg(\dfrac{-15(1-K)^2}{M_1(a)}+\dfrac{8\eta^2\,M_1(a)}{(\eta^2M_5(a)-3b(1-K)\sqrt{-M_5(a)/3})}+4K-7\bigg) \\
M_3(a)=\dfrac{-8\pi\,\eta^2\,a^2\,M_1(a)}{M_4(a)(1-K)^4}\bigg(\dfrac{2\eta^2\,M_4(a)+\sqrt{3M_4(a)}}{(\eta^2\,M_4(a)+\sqrt{3M_4(a)})^2}\bigg)-\dfrac{16\pi\,a^2}{(\eta^2M_5(a)-3b(1-K)\sqrt{-M_5(a)/3})}\\
M_4(a)=\dfrac{\,+3K(K-1)-8\pi\,\rho_{0}\,K\,a^2}{(1-K)^2},\,\,\,M_5(a)=\dfrac{3K(1-K)-8\pi\,\rho_{0}\,K\,a^2}{(1-K)^2} $\\
\\
\begin{figure}[h]
\begin{center}
\includegraphics[width=6cm]{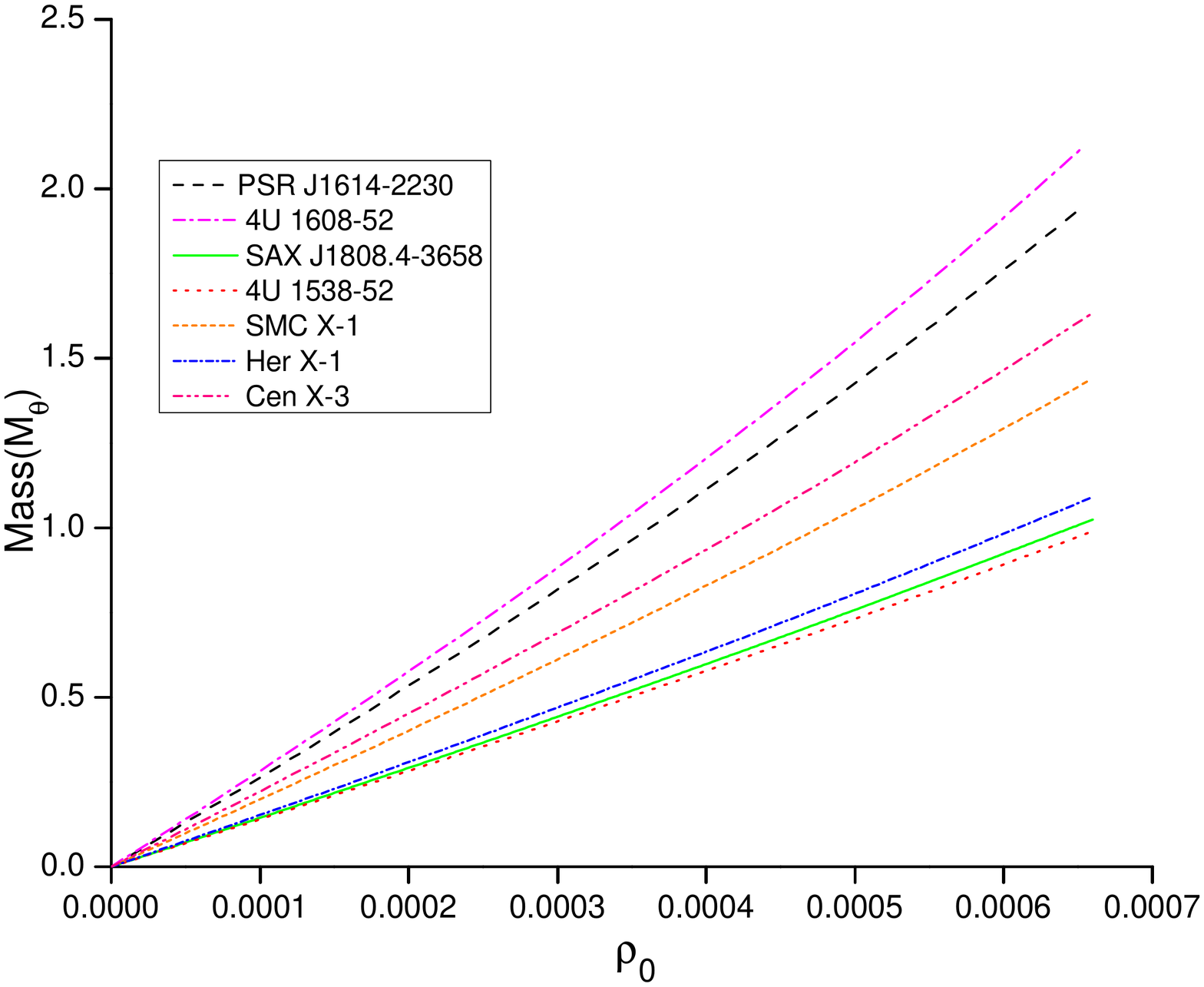}\includegraphics[width=6cm]{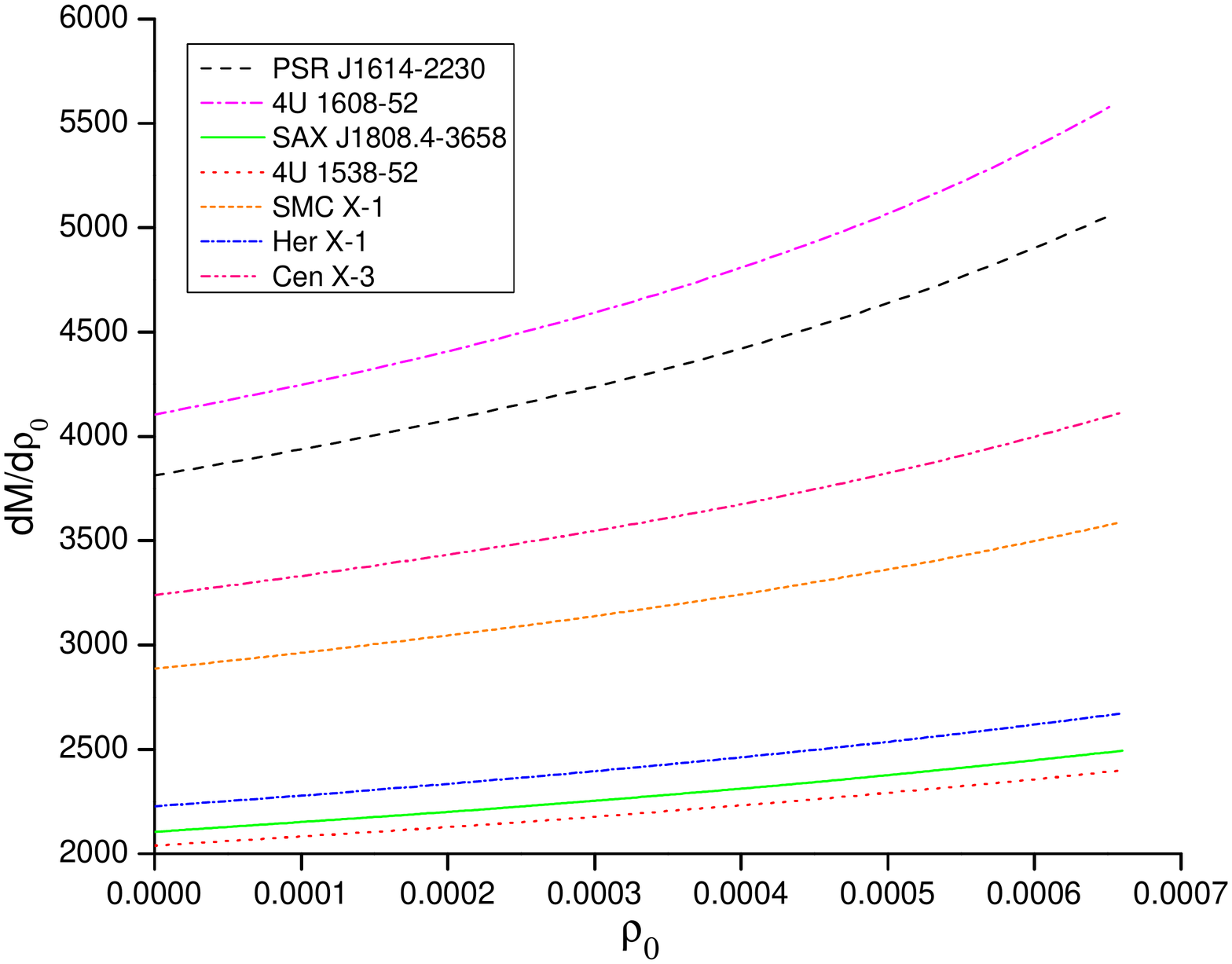}
\caption{Behavior of Mass ($ M_{\odot} $)(left panel) and $ \frac{dM}{d\rho_{0}} $ vs. central density $ \rho_{0} $ for the compact stars PSR J1614-2230,4U 1608-52, SAX J1808.4-3658, 4U 1538-52,SMC X-1,Her X-1 and Cen X-3.In this figure we have used same data set values of physical parameters and constants which are the same in Fig.\ref{f1}} \label{f6}
\end{center}
\end{figure}
From Fig.\ref{f6} we see that the mass $ M(M_{\Theta})$ of isotropic compact star increases with central density $ (\rho_{0}) $. On other hand we observe that $ \dfrac{dM}{d\rho_{0}}  $ is positive with respect to central density $ (\rho_{0}) $. Hence present isotropic compact star model is stable. 
\section{Novelty of Present Model} 
The novelty of our model is that the density$ (\rho) $ is always positive and in monotonically decreasing order for $0<K<1$. If we remove the electric intensity$(q(r)=0)$ or adopt the anisotropic approach then the density$ (\rho) $ is  increasing towards the boundary, i.e.$ \frac{d\rho}{dr}>0\ $ for $0<K<1$. An anisotropic approach we obtained $ \rho_0 $ and $\frac{d\rho}{dr} $ as follows-\\
 $\rho =\dfrac{C(K-1)(3+Cr^2)}{8\pi\,K(1+Cr^2)^{2}}, \\ \dfrac{d\rho}{dr} = \dfrac{-2\,C^2r(K-1)(5+Cr^2)}{8\pi\,K(1+Cr^2)^3}>0 $
 because $ -2K\,C^2r(K -1)>0\,\,\textrm{for}\,\, 0<K<1\, \textrm{and}-1.1446\times 10^{-7}<C<0 $. Also both term $ (5+Cr^2) $ and $ (1+Cr^2)^3$ are positive. Hence the density gradient is also positive.
 \begin{figure}[h]
 \begin{center}
 \includegraphics[width=6cm]{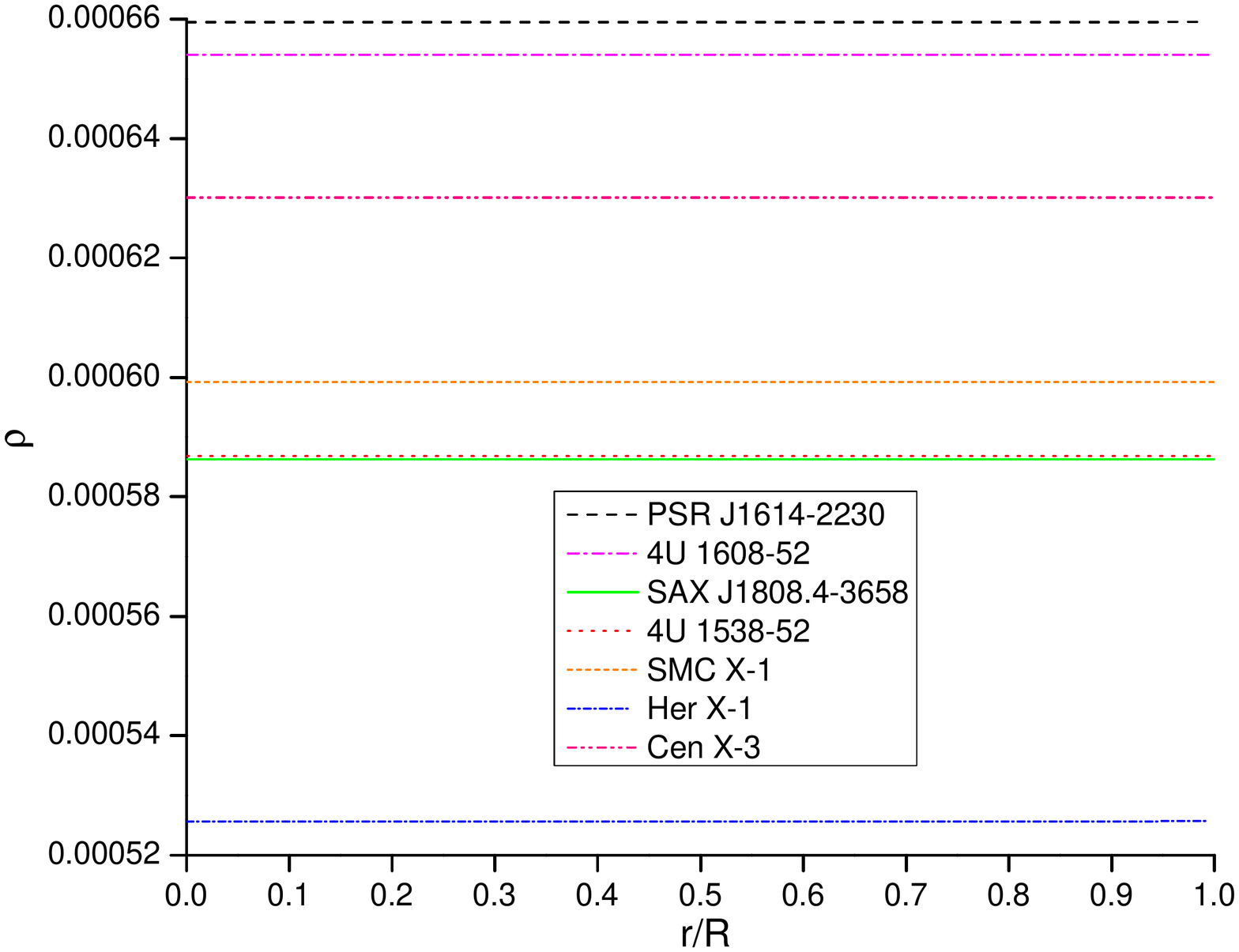} \includegraphics[width=6cm]{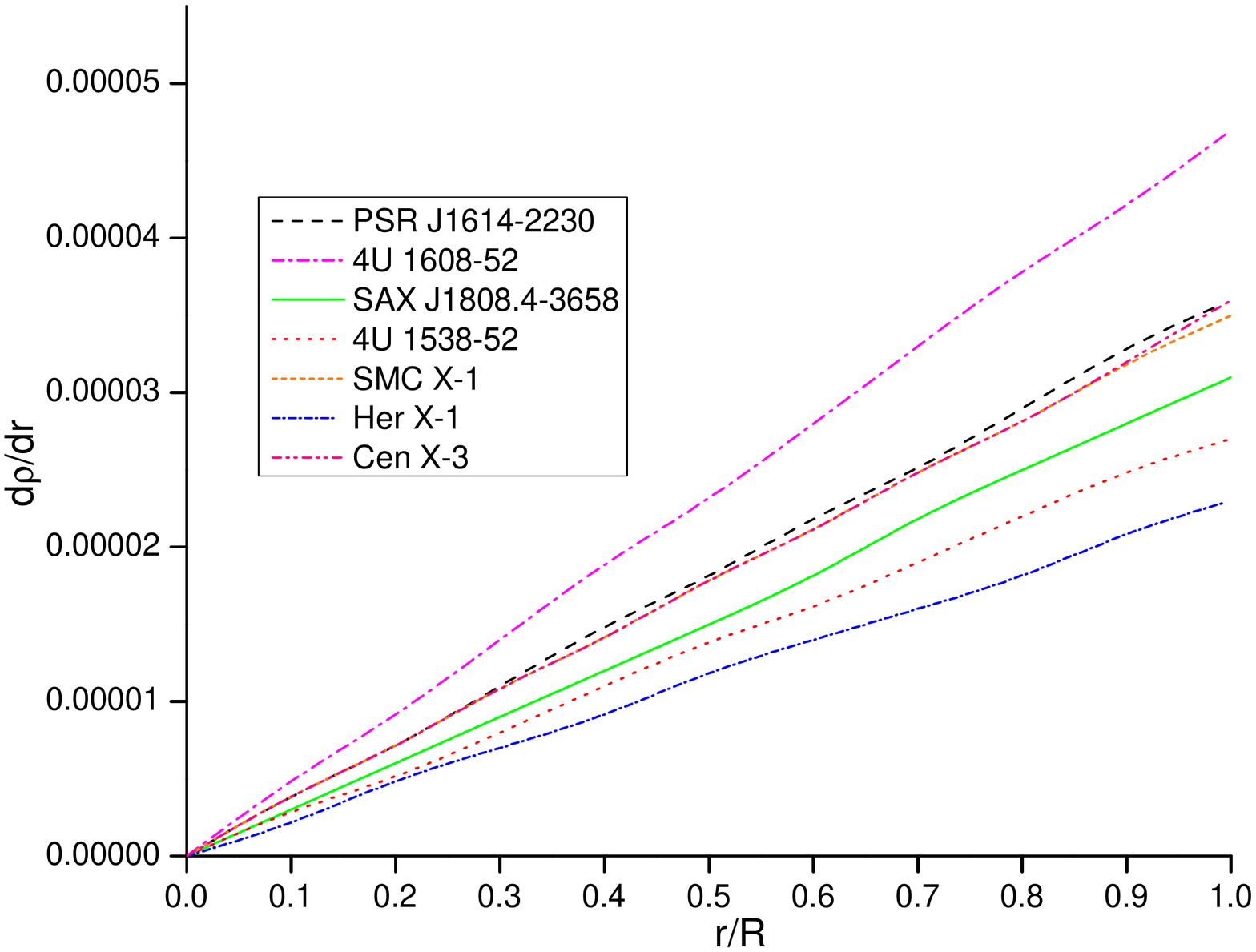}
 \caption{Behaviour of $ \rho_0 $ and $d\rho_0/dr $ vs. fractional radius for PSR J1614-2230, 4U 1608-52  SAX J1808.4-3658,4U 1538-52, SMC X-1,Her X-1 and Cen X-3.In this figure we have used same data set values of physical parameters and constants which are the same in Fig.\ref{f1}}\label{f7}
 \end{center}
\end{figure} 
From Fig.\ref{f7}, we have conclude that our approach is valid with charged compare to anisotropic approach.
\section{Final Remarks}
In this article, we obtained a new class of charged super-dense star models by solving Einstein-Maxwell field equation for a static symmetric distribution of perfect fluid based on a suitable metric potential and considering a particular form of electric intensity.The boundary conditions required for the smooth matching of the interior space-time to the exterior Reissner-Nordstrom space-time which is fixes the constants( see Table \ref{ta1} to \ref{ta8} ) in our solution and determines the mass contained within the charged sphere.In particular, we have demonstrated that the radii and masses measurements for the seven observed compact stars namely,PSR J1614-2230, 4U 1608-52, SAX J1808.4-3658, 4U 1538-52, SMC X-4, Her X-1 and Cen X-3.It could additionally restrict the arbitrary chosen constant parameters and the nature of the stars has been discussed using values of these constants.Graphical analysis of the solution shows that the pressure,density and ratio of $ p/\rho $ are monotonically decreasing towards the surface, when $ K $ lies between $ 0.00001 $ to $0.00004 $. Also we show it in Fig.\ref{f1}-\ref{f5} for $K= 0.0000135, 0.0000209, 0.0000319, 0.0000296, 0.000023, 0.0000277, 0.000018. $ Causality condition is obeyed at each interior point of the configuration. Stability analysis via the Zeldovich stability criterion indicate that our model satisfy causality condition i.e,$(dp/c^{2}d\rho)<1 $ which can be observed from Fig.\ref{f5}(top left). Many authors have discussed causality condition Bludman and Ruderman\cite{1968}\cite{1970}; Krostscheck and Kundt\cite{1978}; Caporaso and Brecher\cite{1979}\cite{g}; Glass\cite{1983}; Morris and Thron\cite{1988}, Curits\cite{1950}. The model is satisfy the following energy conditions(see Fig.\ref{f3}) (i) strong energy condition(SEC),(ii) weak energy condition(WEC) and (iii) null energy condition(NEC).Fig.\ref{f5}( bottom right), we observe that the redshift is also decreasing from the center to surface for $ K$as above.The stability of the charged fluid models depends on the adiabatic index $ \gamma $. Heintzmann and Hillebrandt \cite{heint} proposed that a neutron star model with EOS is stable if $ \gamma >1 $.In this model adiabatic index for $ K $(as above) is greater than $ 4/3 $ see Fig.\ref{f5}(bottom left).The model is also in static equilibrium by solving TOV equation. Table(\ref{ta1}-\ref{ta8}) contain numerical value of physical quantities ,where we used various symbols as follows:\\
$ Zo= $ redshift at the center,  $ Za= $ redshift at the surface,solar mass $ M_{\odot}=1.475km , ~~ G=6.673\times 10^{-8}cm^{3}/gs^{2},~~ c=2.997\times 10^{10}cm/s. ~~$\\

\end{document}